\documentclass[fleqn,usenatbib]{mnras}

\usepackage[T1]{fontenc}
\usepackage{ae,aecompl}

\usepackage{epsf,palatino,url,wrapfig,array,setspace,amsmath,amssymb,fancyhdr,multirow,lscape,appendix,rotating}
\usepackage{graphics,epsfig,txfonts}
\usepackage{graphicx}

\usepackage{longtable}
\usepackage{amssymb}
\usepackage{xcolor}
\usepackage{color}
\usepackage{lipsum}
\usepackage{textcomp}
\usepackage{threeparttable,booktabs}
\usepackage{enumerate}
\usepackage{placeins}
\usepackage{float}
\usepackage[normalem]{ulem}
\usepackage{soul}
\usepackage{blindtext}
\usepackage{bm}

\usepackage{tikz,hyperref}
\newcommand{\ergs}[1]{$\times 10^{#1}$ erg s$^{-1}$}
\newcommand{\oergs}[1]{$10^{#1}$ erg s$^{-1}$}

\newcommand{\ltsima}{$\buildrel < \over \sim$}
\newcommand{\lsim}{\lower.5ex\hbox{\ltsima}}
\newcommand{\gtsima}{$\buildrel > \over \sim$}
\newcommand{\gsim}{\lower.5ex\hbox{\gtsima}}

\newcommand{\swift}{{\it Swift}\xspace}

\newcommand{\xmm}{{\it XMM-Newton}\xspace}

\newcommand{\nicer}{\textit{NICER}\xspace}

\newcommand{\fermi}{{\it Fermi}\xspace}
\newcommand{\nustar}{{\it NuSTAR}\xspace}
\newcommand{\astrosat}{{\it AstroSat}\xspace}

\newcommand{\lxp}{\hbox{RX\,J0520.5-6932}\xspace}
\newcommand{\lxps}{\hbox{RX\,J0520}\xspace}

\newcommand{\rxj}{\hbox{RX\,J0209-7427}\xspace}
\newcommand{\rxjs}{\hbox{RX\,J0209}\xspace}
\newcommand{\swj}{\hbox{Swift\,J0243.6+6124}\xspace}
\newcommand{\swjs}{\hbox{J0243}\xspace}

\graphicspath{{./}{plots2/}}

\DeclareRobustCommand{\VAN}[3]{#2}

\defcitealias{1979ApJ...234..296G}{GL79}
\defcitealias{1995ApJ...449L.153W}{W95}
\defcitealias{2014MNRAS.437.3664H}{H14}

\title[Super-Eddington outbursts torque modelling]{A Bayesian approach for torque modelling of BeXRB pulsars with application to super-Eddington accretors}

\author[Karaferias~A.~S. et al.]
{A.~S.~Karaferias,$^1$
G. Vasilopoulos,$^2$\thanks{E-mail: georgios.vasilopoulos@astro.unistra.fr}
M. Petropoulou,$^1$
P.~A.~Jenke,$^{3}$\newauthor
C.~A.~Wilson-Hodge,$^{4}$
C.~Malacaria$^{4,5}$
\\
$^{1}$Department of Physics, National and Kapodistrian University of Athens, University Campus Zografos, GR 15783, Athens, Greece \\
$^2$Universit\'e de Strasbourg, CNRS, Observatoire astronomique de Strasbourg, UMR 7550, F-67000 Strasbourg, France\\
$^3$University of Alabama in Huntsville, Huntsville, AL 35805, USA\\
$^4$International Space Science Institute (ISSI), Hallerstrasse 6, 3012 Bern, Switzerland\\
$^5$Universities Space Research Association, NSSTC, 320 Sparkman Drive, Huntsville, AL 35805, USA
}

\date{Accepted XXX. Received YYY; in original form ZZZ}

\pubyear{2020}

\begin{document}
\label{firstpage}
\pagerange{\pageref{firstpage}--\pageref{lastpage}}
\maketitle

\begin{abstract}
In this study we present a method to estimate posterior distributions for standard accretion torque model parameters and binary orbital parameters for X-ray binaries using a nested sampling algorithm for Bayesian Parameter Estimation. We study the spin evolution of two Be X-ray binary systems in the Magellanic Clouds, \lxp and \rxj, during major outbursts, in which they surpassed the Eddington-limit. Moreover, we apply our method to the recently discovered \swj; the only known Galactic pulsating ultra-luminous X-ray source. This is an excellent candidate for studying the disc evolution at super-Eddington accretion rates, for its luminosity span several orders of magnitude during its outburst, with a maximum $L_{\rm X}$ that exceeded the Eddington limit by a factor of $\sim 10$. Our method, when applied to \lxp and \rxj, is able to  identify the more favourable torque model for each system, while yielding meaningful ranges for the NS and orbital parameters.
Our analysis for \swj illustrates that, contrary to the standard torque model predictions, the magnetospheric radius ($R_{\rm m}$) and the Alfv\'en radius ($R_{\rm A}$) are not proportional to each other when surpassing the Eddington limit. 
Reported distance estimates of this source
range between 5 and 7 kpc. 
Smaller distances require non-typical neutron star properties (i.e. mass and radius) and possibly lower radiative efficiency of the accretion column.

\end{abstract}




\section{Introduction}

X-ray pulsars (XRPs) are astronomical objects powered by accretion that display periodic variations in X-ray intensity. They are formed when highly magnetized ($B>10^{11}$\,G) neutron stars (NSs) are found in close binary systems, allowing material to be transferred by the donor star onto the NS surface. The spin period evolution of a NS is indicative of the type of its accretion mechanism, the accretion disc structure and the neutron star magnetic field $B$. XRPs can also be useful in developing our understanding of the evolution process of  binary systems with neutron-star members \citep{1997ApJS..113..367B}.

The majority of XRPs are found in Be X-ray binaries (BeXRBs) \citep[see][for a review on BeXRBs]{2011Ap&SS.332....1R}. In this case material escapes the massive donor through a slow equatorial outflow, which is usually known as the decretion disc or the Be disc \citep[e.g.][]{2011A&A...527A..84K}. The mechanism behind the formation and depletion of the Be disc is still a matter of debate, however its transient nature results in highly variable mass transfer, and causes outbursts in BeXRBs. Observations of BeXRBs point to transient activity that is manifested in the form of two types of outbursts \citep[e.g.][]{1986ApJ...308..669S,1997ApJS..113..367B}. Type I outbursts ($L_{\rm X}\sim$\oergs{36-37}) may occur during a close passage of the NS to the decretion disc and thus show a correlation with the binary orbital period. Giant or type II outbursts ($L_{\rm X}\ge$\oergs{38}) that can last for several orbits are associated with warped Be-discs \citep{2013PASJ...65...41O}. 

During outbursts an accretion disc is formed around the NS, resulting in angular momentum transfer to the NS and a change of its spin.  
At a zeroth order approximation the NS spin changes due to mass accretion. However, the consensus is that torque acts through what is called a ``magnetically threaded disc model'' \citep[first introduced by][]{1979ApJ...234..296G} that describes the coupling of the NS magnetic field lines and the accretion disk resulting in torques acting on the NS \citep[see overview by][]{2016ApJ...822...33P}.

Studies of the brightest type II outbursts ($L_{\rm X}>10^{38}$~erg s$^{-1}$) became especially relevant in the advent of the recent discoveries of pulsating ultra-luminous X-ray sources \citep[PULXs, e.g.][]{2014Natur.514..202B,2017Sci...355..817I,2018MNRAS.476L..45C}. ULXs are extragalactic point sources with an apparent isotropic luminosity above the Eddington limit for a 20\,$M_{\odot}$ black hole \citep[see][for a recent review]{2017ARA&A..55..303K}. 
The discovery and subsequent study of PULXs confirmed that at least some ULXs are powered by highly magnetised NSs.
Indeed, an increasing number of authors have put forward the hypothesis that a large fraction of ULXs may actually be powered by strongly magnetized NSs~\citep[see e.g.][]{2017A&A...608A..47K,2017MNRAS.468L..59K,2018ApJ...856..128W}, building upon the early models for XRPs \citep{1976MNRAS.175..395B} but also on more recent theoretical considerations \citep{2015MNRAS.447.1847M}.

Estimates of the NS magnetic field can be made directly through the detection of cyclotron emission lines in XRP spectra \citep{2019A&A...622A..61S}. These lines may be directly formed in the accretion column \citep[e.g.][]{1976MNRAS.175..395B,2015MNRAS.454.2714M} or through reflection onto the NS atmosphere \citep[e.g.][]{2013ApJ...777..115P,2021arXiv210807573K}. However, these direct measurements are hindered by the spectral resolution and energy range of our instruments that make difficult the detection of lines corresponding to magnetic field strengths $B\gtrsim 10^{13}$~G. Note however that \textit{INTEGRAL} has pushed this limit for nearby bright systems  \citep{2003A&A...411L...1W}.
Alternatively, indirect measurements of $B$ may be derived from the spin evolution of the NS during major outbursts. Such calculations require the use of torque models and proper corrections for the orbital motion of the binary. This method may be applied to systems with a wide range of magnetic field strengths, including PULXs.

Given that PULXs host magnetized NSs several authors have invoked standard torque models \citep[i.e.][]{1979ApJ...234..296G,1995ApJ...449L.153W} to estimate the magnetic field of the NS 
\citep[e.g.][]{2018A&A...620L..12V,2019MNRAS.488.5225V,2020MNRAS.491.4949V,2020ApJ...891...44B,2020ApJ...899...97E,2021JHEAp..31....1C}. At the same time theoretical studies have demonstrated that it is required to adjust these standard torque models due to change in the disc structure when exceeding the Eddington limit \citep[e.g.][]{2009A&A...493..809B}. Moreover, according to \citet{2017MNRAS.470.2799C,2019A&A...626A..18C}, and their numerical calculations, the radius of the magnetosphere should not be regarded as being to scale with the Alfv\'en radius for all mass accretion rates as suggested by the standard models \citep{1977ApJ...217..578G, 1991ApJ...370L..39K, 1996ApJ...465L.111W, 2007ApJ...671.1990K}. Instead, the ratio of the magnetospheric and Alfv\'en radii was found to depend on the mass accretion rate in a way that leads to an almost constant magnetospheric radius for super-Eddington mass accretion rates \citep[see also][]{2019MNRAS.484..687M}.

In this work we study the spin evolution of accreting NSs during major outbursts of BeXRBs that reached or exceeded the Eddington limit using torque models that are widely used in the literature \citep[i.e.][]{1979ApJ...234..296G,1995ApJ...449L.153W, 2014MNRAS.437.3664H}. We also implement a nested sampling algorithm for Bayesian parameter estimation and apply it to our sample of sources to simultaneously estimate posterior distributions for the parameters of standard accretion torque models and binary orbital parameters.

We first test our approach against \lxp (\lxps hereafter) and data obtained during a major outburst in 2014 that lasted for several orbits. 
Then we apply our method to two of the most energetic systems monitored by the \fermi Gamma-ray Burst Monitor \citep[GBM,][]{2009ApJ...702..791M},
namely \rxj (\rxjs hereafter) and \swj (\swjs hereafter).
For \lxps and \rxjs we found that our method converges to a solution with almost no fine-tuning of the parameter space. In addition it provides more realistic uncertainties to the model parameters than typical methods based on least square fitting, and also enables investigation of degeneracies between parameters.  
The challenge was the modelling of \swjs, a system with data that cover a large dynamic range in luminosity, and with maximum luminosity exceeding the Eddington limit by a factor of $\sim$ 10 considering a distance of $\sim$ 5-7~kpc \citep{2020A&A...640A..35R, 2018A&A...613A..19D}.
The big variation of the bolometric luminosity of \swjs provides us with an excellent test case to examine the relation of the magnetospheric radius with the accretion rate. Standard accretion models for \swjs should 
be modified to account for the change in the magnetospheric radius at super-Eddington accretion rates, as demonstrated in recent theoretical and observational studies \citep[e.g.][]{2017MNRAS.470.2799C,2019A&A...626A..18C,2019MNRAS.484..687M,2019A&A...626A.106M,2020MNRAS.491.1857D}. For this purpose, we find a parametric expression of the coefficient $\xi$, which is defined as the ratio of the magnetospheric radius to the Alfv{\'e}n radius, as a function of the accretion rate. In other words, we move beyond the assumption of a constant $\xi$, usually made in the study of accreting pulsars. Our empirical approach would be applied for the first time in observational data of systems above the Eddington limit, but we refer the reader to \citet{2009A&A...493..809B} for a parametric study of the disc-magnetosphere interaction models in lower luminosity accreting systems. 

This paper is structured as follows. In Secs.~\ref{sec:models} and \ref{sec:obsdata} we outline respectively the torque models and the observational data that will be used in our study. In Sec.~\ref{sec:method} we present our methodology for modelling the spin evolution of the XRPs in our sample, and describe the Bayesian approach we implemented for the latter. In Sec.~\ref{sec:application} we introduce the three systems in our sample and present the results of our analysis for each source in Sec.~\ref{sec:results}. We continue with a discussion of our results in Sec.~\ref{sec:dis} and finish with our conclusions in Sec.~\ref{sec:conclusion}.

\section{Accretion torque models}\label{sec:models}

The problem of mass and torque transfer in accreting NS has been investigated by several studies in the past 50 years \citep[e.g. see][and references within]{2002apa..book.....F,2016ApJ...822...33P}. In the following paragraphs we will introduce the basic equations that we invoked in our work.

Assuming spherical accretion, the gas will stop at the so-called Alfv\'{e}n radius, which is estimated by equating the magnetic pressure from the stellar dipole to the ram pressure of gas free-falling from infinity \citep{1977ApJ...215..897E, 1973ApJ...179..585D}:
\begin{equation}
R_{\rm A} = \left(\frac{\mu^{4}}{2GM \dot{M}^{2}}\right)^{1/7}, 
\label{Ra}
\end{equation}
where $M$ is the NS mass, $\mu=B R^{3}$ is the magnetic dipole moment, with $R$ the NS radius and $B$ the NS magnetic field strength at the equator\footnote{Alternatively $\mu=B_{\rm p} R^{3}/2$, where $B_{\rm p}$ is the field at the magnetic poles as opposed to the equator.}, $\dot{M}$ is the accretion rate, and $G$ is the gravitational constant.

We define the truncation radius of a thin Keplerian disc as the magnetospheric radius
\begin{equation}
R_{\rm m}=\xi R_{\rm A},
\label{Rm}
\end{equation}
where $\xi\sim 0.5\:-\:1$ for all kinds of magnetic stars \citep[see][]{2018A&A...610A..46C}.

After material gets halted at $R_{\rm m}$ it may continue flowing towards the NS if its angular momentum is high enough to penetrate the centrifugal barrier set by the rotating magnetosphere. The radius where a particle attached to a field line would rotate at the Keplerian rate is defined as the corotation radius and is expressed as:
\begin{equation}
R_{\rm co} = \left(\frac{GM}{\Omega^{2}}\right)^{1/3},
\end{equation}  
where $\Omega$ is the NS angular velocity.
Since matter inside the corotation radius flows along the field lines, for steady accretion to occur, the Keplerian angular velocity at $R_{\rm m}$ has to be larger than the angular velocity of the star (and of the field lines).
Following this rational \citet{1977ApJ...215..897E} defined the fastness parameter as:
\begin{equation}
\omega_{\rm fast} = \frac{\Omega}{\Omega_{\rm K}(R_{\rm m})} = \left(\frac{R_{\rm m}}{R_{\rm co}}\right)^{3/2},
\end{equation} 
where $\Omega_{\rm K}(r)$ is the Keplerian angular velocity at distance $r$.

A major consequence of accretion and the general interaction of the disc with the NS through its field lines is that the NS spin can change as a result of the induced torques \citep[e.g.][]{1979ApJ...234..296G}.
On the one hand, there is the torque applied to the star by the accretion, $N_{\rm acc}$, defined as
\begin{equation}\label{Nacc} 
N_{\rm acc} = \dot{M}\sqrt{G M R_{\rm m}}.
\end{equation} 
On the other hand, there is a torque, $N_{\rm field}$ that tends to spin-down the pulsar, and is  applied by the dragging of the field lines by the disc and the sweeping of the open field lines due to the effective inertia of the electromagnetic field \citep[see e.g.][]{1997A&A...327..662B}.
The total torque $N_{\rm tot}$ is the sum of the two terms and it is usually expressed as
\begin{equation}\label{Ntot} 
N_{\rm tot} = n\left(\omega_{\rm fast}\right)N_{\rm acc}, 
\end{equation} where $n(\omega_{\rm fast})$ is a  function of the dimensionless fastness parameter that incorporates the details of $N_{\rm field}$ \citep{2016ApJ...822...33P}. 

In the literature several torque models have been developed to explain the coupling of the disc with the magnetosphere and to estimate the induced torque onto the NS \citep[e.g.][]{1979ApJ...234..296G,1995ApJ...449L.153W,2007ApJ...671.1990K,1995MNRAS.275..244L,2004ApJ...606..436R}. In our study we will focus on the \citet{1979ApJ...234..296G} model (hereafter GL79) and the \citet{1995ApJ...449L.153W} model (hereafter W95), as they are the most commonly used in the literature for accreting pulsars.

\citetalias{1979ApJ...234..296G} proposed that the dimensionless function of the fastness parameter may be expressed as
\begin{equation}\label{N_GL79} 
n(\omega_{\rm fast}) = 1.39 \dfrac{1 - \omega_{\rm fast} \left[4.03(1-\omega_{\rm fast})^{0.173}-0.878\right]}{1-\omega_{\rm fast}}.
\end{equation}

\citetalias{1995ApJ...449L.153W} argued for a different 
toroidal magnetic structure than \citetalias{1979ApJ...234..296G}, and recalculated the dimensionless function, which reads
\begin{equation}\label{N_W95} 
n(\omega_{\rm fast}) = \dfrac{7/6 - (4/3)\omega_{\rm fast} + (1/9)\omega_{\rm fast}^{2}}{1-\omega_{\rm fast}}.
\end{equation}
These models, expressed as seen in Eqs.~(\ref{Ntot})-(\ref{N_W95}) are only applicable when $\omega_{\rm fast} < 1$. When we study bright systems during outbursts, we tend to ignore the $\omega_{\rm fast}$ terms in these expressions, as generally $\omega_{\rm fast}\ll1$; in other words, the systems are away from equilibrium - a state in which the NS rotation frequency is constant. However, during the low luminosity phases the assumption of $\omega_{\rm fast}\ll1$ might not hold and the full version of Eqs.~(\ref{N_GL79}) and (\ref{N_W95}) should be used for the treatment of the torque evolution.

When we have transitions from $\omega_{\rm fast} < 1$ to $\omega_{\rm fast} > 1$ (i.e. the system goes through equilibrium) the \citetalias{1979ApJ...234..296G} and \citetalias{1995ApJ...449L.153W} models can no longer be applied. \swjs is such an example. To model the spin evolution in such systems we can use an approximate expression for the total torque that reads
\begin{equation}\label{N_WT20} 
N_{\rm tot} = \dot{M} R_{\rm m}^{2} \Omega_{\rm K}(R_{\rm m})\left(1-\frac{\Omega}{\Omega_{\rm K}(R_{\rm m})}\right).
\end{equation}
Even though several studies have used the approximation of Eq.~(\ref{N_WT20}) \citep[e.g.][]{1999ApJ...520..276M, 2020MNRAS.492..762W}, \cite{2014MNRAS.437.3664H}  have perhaps presented the first extended application to accreting XRPs for the study of their equilibrium state.
In what follows, we therefore refer to Eq.~(\ref{N_WT20}) as the \citetalias{2014MNRAS.437.3664H} model.
Finally, the equation describing the spin up of the NS is given by
\begin{equation}\label{dotv} 
\dot{\nu} \equiv \frac{\dot{\Omega}}{2\pi}= \frac{N_{\rm  tot}}{2 \pi I}
\end{equation}
where $I$ is the NS moment of inertia and $N_{\rm tot}$ may be derived by Eqs.~(\ref{Nacc}) and (\ref{N_GL79}), (\ref{N_W95})  or (\ref{N_WT20}). 
Using the above prescription one may indirectly estimate one of the fundamental parameters of the NS, its magnetic field strength. This is made possible because the $\dot{\nu}$ and $\dot{M}$ can be inferred from observations, while parameters like the NS mass and radius are well determined.

\section{Observational data}\label{sec:obsdata}
Our methodology requires measurements of the spin period and mass accretion rates during major outbursts. It is crucial to obtain a baseline of measurements that would allow an estimation of the orbital parameters and the intrinsic spin-up due to accretion. 

Outbursts are daily monitored in the X-rays by all-sky surveys like the \swift Burst Alert Telescope \citep{2005SSRv..120..143B} (BAT, 15-150 keV), the \fermi Gamma-ray Burst Monitor \citep{2009ApJ...702..791M} (GBM, 8-40 keV) and the Monitor of All-sky X-ray Image \citep{2000AIPC..504..181M} (MAXI, 0.5-30 keV). Moreover pointing observations may be performed by various observatories. In particular, the Neutron star Interior Composition Explorer (\nicer) \citep{2016SPIE.9905E..1HG} and the \swift X-Ray Telescope (XRT) \citep{2005SSRv..120..165B} can perform multiple short observations (i.e. 1-2 ks) over weeks or months; thus, they are ideal for monitoring systems in the soft X-rays (i.e. 0.2-10~keV). Target of opportunity observations  may also be performed by the Nuclear Spectroscopic Telescope Array \nustar \citep{2010HEAD...11.4601H} (3-79 keV) or \astrosat \citep{2014SPIE.9144E..1SS} (0.3-100~keV). These triggered observations last typically over 20 ks and are not repeated more than a couple times over the course of a major outburst. 
However, they are crucial as they deliver broadband spectra with high energy resolution and enable proper characterization of the spectral shape, the bolometric luminosity and the mass accretion rate.  

In our study we will mainly use results that are available in the literature (through repositories), and perform limited data reduction of \swift/XRT data. For the latter case, we retrieved and analyzed the data from the UK \swift \ science data centre\footnote{\url{http://www.swift.ac.uk/user_objects/}} using standard procedures as outlined in \citet{2007A&A...469..379E,2009MNRAS.397.1177E}. 

\subsection{Spin-period monitoring} 
While spin period measurements may be obtained from monitoring observations by \nicer or \swift/XRT, 
they usually have larger uncertainties than the \fermi/GBM measurements. Hence, in this work, we will use \fermi/GBM data products from the GBM accreting pulsar project\footnote{\url{https://gammaray.msfc.nasa.gov/gbm/science/pulsars.html}} to study the spin evolution of the NS \citep[for details see][]{2020ApJ...896...90M}. These products contain spin measurements of data chunks that are typically binned every one to three days depending on the source luminosity.

\subsection{Mass accretion rate estimation}\label{sec:mdot}
As it was discussed in Sec. \ref{sec:models}, measuring the mass accretion rate through monitoring observations is crucial for the torque modelling. There are several ways one can deduct $\dot{M}$ from observational proxies.

The \fermi/GBM products contain pulsed fluxes for epochs where a spin period could be obtained. Pulsed fluxes have been known to correlate with the luminosity of the pulsar, and even $\dot{\nu}$ \citep[e.g. for 2S 1417-624][]{1996A&AS..120C.209F}. 
However, pulsed fluxes are affected by changes in the pulse profile and the pulse shape. Nevertheless, in some systems, like \rxj,  changes in the pulse profile are minimal; \citet{2020MNRAS.494.5350V} showed that the pulse shape remained almost constant during the evolution of its 2019 outburst.
Thus, in certain cases the pulsed flux could be a good proxy of the accretion rate.

Alternatively one could use the \swift/BAT transient monitor results
provided by the \swift/BAT team\footnote{\url{https://swift.gsfc.nasa.gov/results/transients/weak/}} that delivers daily binned data products in the form of count rates \citep{2013ApJS..209...14K}. The advantage of BAT over GBM is that it provides intensities that are not tied to pulsation searches. Nevertheless, BAT count rates have larger uncertainties and more scatter (i.e. day-to-day) compared to the GBM detection of the same source. Thus, it is often required to bin the data over longer intervals, perform some smoothing of the overall light curve and even exclude outliers that often appear as flaring or dipping points\footnote{The \swift/BAT team advises that large positive (or negative) fluctuations for a source on a single day, should be treated with caution, as they are likely not physical.}.

Upon selecting a proxy for the intensity, the next crucial step is the conversion to bolometric X-ray luminosity and finally $\dot{M}$. Ideally, for this purpose one should use broadband spectra (i.e. 0.3-70 keV). These can be obtained using a combination of instruments like \nustar and \astrosat for the hard band ($\sim$1-100 keV) with \xmm, \swift or \nicer for the soft band ($\sim$0.2-10 keV). The absorption-corrected flux can be estimated through spectral fitting, and can be converted to bolometric X-ray luminosity, $L_{\rm X}$. The latter can be then translated to $\dot{M}$ adopting some efficiency $\eta_{\rm eff}$
under which gravitational energy is converted to radiation \citep[typically assumed to be 100 per cent,][]{2018A&A...610A..46C}, namely $L_{ \rm X}=G\dot{M}M/R \approx 0.2~\dot{M} c^2 (M/1.4 M_\odot)(R/10~{\rm km})^{-1}$ \citep{2002apa..book.....F}.


\section{Methodology and implementation}\label{sec:method}

In this section we will discuss the methodology we used when applying our model to the observed data of accreting pulsars.
We will present the model parameters and the steps followed to construct the model. For the application to the data we will employ a Bayesian approach to ensure accurately estimated model parameters and their associated uncertainties.

\subsection{Modelling the intrinsic spin-up}

The first step is to estimate the intrinsic spin-up based on the selected torque model. The free parameters of the model are: 
\begin{itemize}
    \item the magnetic field strength at the NS equator $B$,
    \item the ratio of the magnetospheric radius to the Alfv{\'e}n radius (i.e. the $\xi$ parameter),
    \item the NS spin frequency at some reference time (i.e. $v_{\rm 0}$) and, 
    \item the distance $d$ to the source. 
\end{itemize}

In general there is a degeneracy between $\xi$, $B$ and $d$. 
For example, if the system is away from equilibrium with $\omega_{\rm fast} \ll 1$ there is a power-law dependence\footnote{This is derived by the standard power-law dependence usually taken as $B \propto (PL_{\rm X})^{-6/7}$} of $B$ on $d$ (i.e. $B \propto d^{-6/7}$).
This scaling can be easily understood as follows.
For a larger distance $d$ (and the same observed flux), the derived $L_{\rm X}$ and $\dot{M}$ are higher, thus a lower $B$ field is needed to explain the measured $\dot{\nu}$. However, a lower magnetic field means that the spin-equilibrium is going to be reached at lower fluxes. 
If the transition to spin-equilibrium is covered by monitoring observations, the degeneracy may be partially broken, 
since we now need to make assumptions for only one of the three free parameters $\xi$, $B$ and $d$.  The above discussion motivates studies of extragalactic XRBs where the distance of the host galaxies is well determined, such as the Magellanic Clouds. A similar power-law dependence exists between $B$ and $\xi$. For thin accretion discs one may restrict $\xi$ within a small range of values and in most cases it is safe to assume $\xi=0.5$ \citep{1977ApJ...217..578G}. Nevertheless, we will also consider a mass-accretion dependent $\xi$ parameter whenever relevant (see \ref{sec:ksi}).

To calculate the intrinsic spin-up of the NS as a function of time, $\nu(t)$, for the three torque models described in Sec. \ref{sec:models}, we used Eq.~(\ref{dotv}) and the inferred $\dot{M}(t)$ from one of the proxies described in Sec. \ref{sec:mdot}. To minimize any sawtooth-like effects in the derived time series of the mass accretion rate, we re-sampled it with a finer resolution (i.e. 10-20 steps per day). 

In all calculations we assumed  a typical value for the NS moment of inertia, i.e. $I = 1.3\times10^{45}$~g~cm$^{2}$ unless stated otherwise, and considered that $R_{\rm co}$ is constant in time. While the corotation radius can in principle evolve during an outburst, the expected change is very small given the minimal change in ${v}$ and the large dynamical range of $L_{\rm X}$ that drives the $\dot{\nu}$ evolution during an outburst (see Sec.~\ref{sec:application}). 

\begin{figure*}
    \centering
    \includegraphics[width=\textwidth]{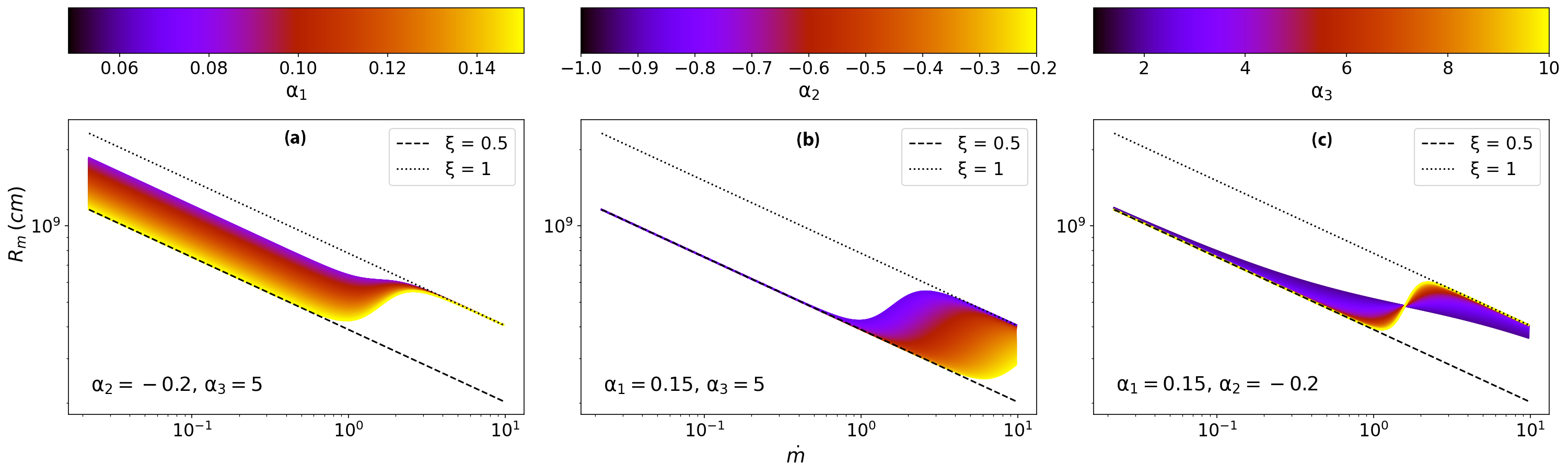}
    \caption{
    The magnetospheric radius given by Eq.~(\ref{eq:Rm}) plotted against the dimensionless accretion rate for different choices of the free parameters $a_{\rm 1}$, $a_{\rm 2}$ and $a_{\rm 3}$ (from left to right). Other parameters used are: $a_{\rm 0}=3.459$, corresponding to $B = 10^{13}$~G for $M=1.4 M_{\sun}$ and $R=1.2\times10^6$~cm.
    Each $a_{\rm 1}$ value translates to a different $\xi$ range and we expect $\xi_{\min} \gtrsim 0.5$ for  $a_{\rm 1}\lesssim 0.14$. The $a_{\rm 2}$ parameter indicates the characteristic value of  $\dot{m}$ where the outflows become significant and therefore the $\xi$ parameter deviates from the standard $\sim0.5$ value. The $a_{\rm 3}$ parameter determines how wide is the area where $\xi$ deviates from a constant value.}
    \label{fig:model_test}
\end{figure*}

\subsection{A mass accretion dependent $\xi$}\label{sec:ksi}

The value of the $\xi$ parameter is motivated by theory of disc accretion \citep[e.g.][]{1977ApJ...217..578G}. 
The inner region of a geometrically thin disc is gas-pressure dominated and $\xi{\approx}0.5$ \citep{2018A&A...610A..46C}. 
However, as the accretion rate increases, radiation pressure becomes increasingly more important and the disc structure changes (becoming geometrically thick). Significant outflows from the disc play also an important role in this picture. 
It has been shown \citep[e.g. Eq. 61 of][]{2017MNRAS.470.2799C} that at a given mass accretion rate the magnetospheric radius becomes almost independent of the accretion rate, if the radiation pressure dominates at the inner parts of a disc \citep[see also][]{2019MNRAS.484..687M}. In this case the $\xi$ parameter gradually changes from $\sim$0.5 to $\sim$1.0 as a function of mass accretion rate. For even higher accretion rates the advection of viscously generated heat in the inner disc plays a more important role. Because of heat advection the radiation energy flux transported by diffusion in the vertical direction is less than the one released locally in the disc. As a result, the advection process effectively leads to a reduced mass loss from the disc~\citep[e.g.][]{2019MNRAS.484..687M} and the
relation between the magnetospheric radius and the accretion rate is closer to that of the standard models (i.e. $R_{\rm m} \sim \dot{M}^{-2/7}$) \citep[see Fig. 12 of][]{2019A&A...626A..18C}.
This change in disc structure is supported also by observational evidence in the power density spectra of pulsars \citep[e.g.][]{2019A&A...626A.106M, 2020MNRAS.491.1857D}.

When modelling the spin evolution of XRPs a constant $\xi$ is usually assumed, as it is rare to observe a transition through the above mentioned accretion regimes during an outburst. 
Although extragalactic BeXRBs have been known to exceed the Eddington limit, it is difficult to obtain quality data at lower luminosity levels.
Thus, \swjs offers a unique case-study with quality monitoring observations spanning over a large dynamical range around the Eddington limit. Therefore, we will implement an accretion-dependent $\xi$ parameter in the modelling of this source, as described below. 

Taking the logarithm of Eq.~(\ref{Rm}) and using Eq.~(\ref{Ra}) we may write
\begin{equation}
\label{eq:Rm}
    \log \frac{R_{\rm m}}{R_{\rm g}} = \log[\xi(\dot{m})] -\frac{2}{7}\log \dot{m} + 
    a_{\rm 0}
\end{equation}
where $a_{\rm 0}$ is defined as
\begin{equation}
    a_{\rm 0} = 
    \log\left[ \frac{1}{R_{\rm g}}\left(\frac{B^{4}R^{12}}{2GM\dot{M}_{\rm Edd}^{2}}\right)^{1/7}\right].
    \label{Bmodel}
\end{equation}
In the equations above $R_{\rm g} = GM/c^{2}$ is  the NS gravitational radius, and $\dot{m}$ is the accretion rate normalized to the Eddington accretion rate for a NS, $\dot{M}_{\rm Edd}$. This is defined from the relation $GM\dot{M}_{\rm Edd}/R = L_{\rm Edd} = 4 \pi c G M/\kappa$, namely $\dot{M}_{\rm Edd} = 4 \pi c R / \kappa \simeq 1.3 \times 10^{18} (R/12~{\rm km})$~g s$^{-1}$, where $\kappa\approx 0.2(1+X)$~cm$^2$ g$^{-1}$ is  the Thomson opacity and $X=0.7$ is the hydrogen abundance. For typical NS parameter values, i.e. $B=10^{12-13}$~G, $R=12$~km, and $M=1.4 M_{\sun}$ we find $a_{\rm 0}\simeq 3 - 3.5$.

Motivated by the results of \citet{2019A&A...626A..18C} we developed a functional form for the $\xi$ parameter, namely
\begin{equation}
\log \xi = a_{\rm 1}\:(\tanh[(\log\dot{m}\: +\: a_{\rm 2})\: a_{\rm 3}] - 1),
\label{xi}
\end{equation}
where 
$a_{\rm 1},a_{\rm 2}$, and $a_{\rm 3}$
are parameters to be determined by the fit to the data (see Sec. \ref{sec:results}). Here, $a_{\rm 1}$ describes the range of $\xi$ values, $a_{\rm 2}$ describes the range of $\dot{m}$ values where $R_{\rm m}$ deviates from the standard scaling relation (i.e. $R_{\rm m} \propto \dot{m}^{-2/7}$), and $a_{\rm 3}$ describes how fast $\xi$ changes. To better illustrate the dependence of $R_{\rm m}$ on $\dot{m}$ we plot Eq.~(\ref{eq:Rm}) in Fig. \ref{fig:model_test}  for different choices of the parameters $a_{\rm 1}, a_{\rm 2}$, and $a_{\rm 3}$. 

\subsection{Modelling orbital spin evolution}\label{sec:orbit}
The Doppler shifts induced by the orbital motion in an XRP are described with five orbital parameters: orbital period ($P_{\rm orb}$), orbital eccentricity (hereafter $e$), the epoch of a mean longitude of 90 degrees of the star's orbit ($T_{\rm \pi/2}$),  
the semi-projected binary separation ($a\sin i$) and the orbital phase that is commonly expressed as the angle of periapse ($\omega$).

Synthesizing NS radial velocities for a set of orbital parameters and at given times involves solving Kepler's equations, which can be done by an iterative method \citep[e.g.][]{1988fcm..book.....D,2018PASP..130d4504F}. Upon computing the radial velocities, one can combine it with the intrinsic spin evolution to derive a complete model for the evolution of the NS frequency in time. 
\begin{equation} 
v_{\rm model} = v\:(1-V_{\rm r}(\boldsymbol{\theta}_{\rm Kep})) \end{equation} 
where $V_{\rm r}$ refers to the radial velocity of the NS due to the Keplerian orbit with parameters contained in the vector $\boldsymbol{\theta}_{\rm Kep}$.

\subsection{Bayesian Inference - {\sc ultranest}}

Our goal is to infer the posterior
probability density $p$ given a dataset ($\mathcal{D}$) and priors from the Bayes' Theorem for a model with a set of parameters contained in the vector $\boldsymbol{\theta}$.
Having calculated the model we construct a likelihood function.
Given the nature of the physical problem we added a term $\ln{f}$ to account for the systematic scatter and noise of our data not included in the statistical uncertainties of the measurements. This term results in an excess variance compared to statistical uncertainties, i.e.
\begin{equation}
\sigma_{\rm tot, \rm i}^2 = \sigma_{\rm i}^2 + e^{2\ln{f}}, 
\end{equation}
where $\sigma_{\rm i}$ are the GBM frequency errors, $\sigma_{\rm tot, i}$ are the errors after accounting for the systematic scatter and noise not included in the statistical uncertainties of the measurements and $i$ runs over the times of measurements.
The likelihood function for a dataset $D_{\rm j}$ can be then written as:
\begin{equation}
\ln{\mathcal{L}_{\rm j}(\boldsymbol{\theta}|\mathcal{D}_{\rm j})} =  -\frac{1}{2} \sum_{i}^{} \frac{(\nu_{\rm model}-\nu_{\rm data})^2}{\sigma_{\rm tot, i}^2}+e^{\sigma_{\rm tot, i}^2}, 
\end{equation}
where $\nu_{\rm data}$ are the measured spin frequencies. In principle, different datasets ($j=1,\cdots,N$) can be combined to construct the total likelihood function of the model. 

To derive the posterior probability distributions and the Bayesian evidence we used the nested sampling Monte Carlo algorithm
MLFriends \citep{2004AIPC..735..395S,2019PASP..131j8005B} that employs the
{\sc ultranest}\footnote{\url{https://johannesbuchner.github.io/UltraNest/}} package \citep{2021JOSS....6.3001B}. 
The overall procedure is similar to methods used to derive Keplerian orbits from the time series of radial velocities \citep[e.g.][]{2018PASP..130d4504F} that use Markov
Chain Monte Carlo (MCMC) methods \citep[][]{2013PASP..125..306F}.
The advantage of using {\sc ultranest} lies in its overall strengths that are the unsupervised navigation of complex, potentially multi-modal posteriors until a well-defined termination point. Thus, no initial optimization is needed and minimal adjustment of the priors is necessary.

\section{Application to systems}\label{sec:application}
We present the systems that will be used as test beds of our methodology. We selected three BeXRBs that underwent outbursts exceeding the Eddington limit and are listed in Table~\ref{tab:systems}. 
 
\begin{table}
	\centering
	\small
    \setlength\tabcolsep{1pt}
	\caption{List of systems considered in this study.}
	\label{tab:systems}
	\begin{tabular}{lccc}
	\hline 
	Name & Distance  &   Outburst epoch  & Observatories$\dagger$ \\  &  (kpc) & (MJD) &   \\  \hline \hline \smallskip
	RX\,J0520.5-6932 & $\sim 50$ & 56645.3-56723.5 &  F, N, SX, Nu,\\
	\hline\smallskip
	RX\,J0209.6-7427 & $\sim 62$  & 58807.0 - 58887.0 & F, N, Nu,\\
	\hline\smallskip
	Swift J0243.6+6124 & $\gtrsim 5$ & 58027.5 - 58497.5 & F, SB, Nu\\
	\hline 
	\end{tabular}
 \begin{tablenotes}\footnotesize
 \item $\dagger$ Observatories whose data we used in this study: \fermi/GBM (F), \nicer (N), \nustar (Nu), \swift/XRT (SX) and BAT (SB).\\ 
 \end{tablenotes}
\end{table}

RX\,J0520 is a BeXRB located in the Large Magellanic Cloud (LMC) hosting a 8.04~s pulsating NS 
\citep[i.e. LXP 8.04][]{2014A&A...567A.129V}. In 2014 the system went through a major outburst that exceeded the Eddington limit \citep{2014ATel.5760....1V,2014ApJ...795..154T}. The 2014 major outburst lasted for several months and was monitored by \fermi/GBM, \swift/XRT and \swift/BAT. \fermi/GBM monitoring resulted in determination of orbital parameters of the system \citep{2020ApJ...896...90M}. 
The major outburst was monitored by \fermi/GBM and \swift/BAT all sky detectors and by pointed \swift/XRT observations for more than seven orbital periods. Given that GBM detected pulsations for about 80 consecutive days, this makes the system an ideal test-case for our method. 

RX\,J0209 is a BeXRB located in the outer wing of the Small Magellanic Cloud (SMC) hosting a 9.3 s pulsating NS \citep{2020MNRAS.494.5350V}. In November 2019 it exhibited a particularly bright outburst, among the brightest we have observed from a BeXRB in the Magellanic Clouds, reaching super-Eddington luminosity, that was detected by MAXI. During the outburst, the system was monitored by \nicer, \fermi/GBM, \astrosat and \swift/BAT. Furthermore \fermi/GBM monitoring resulted in determination of preliminary orbital parameters of the system\footnote{GBM Accreting Pulsars project:\\ \url{https://gammaray.msfc.nasa.gov/gbm/science/pulsars.html}}.

Swift\,J0243 is the first and only known Galactic PULX \citep{2018ApJ...863....9W}. It was first detected by \swift/BAT on October 3, 2017 \citep{2017ATel10809....1K} during an outburst that lasted until 2018. This source is characterized by a spin period of $\sim$9.86 s \citep{2017ATel10812....1J} and at its peak the $L_{\rm X}$ is well above the Eddington limit ($\sim10^{39}$erg/s) \citep{2018A&A...613A..19D}. During the period over which we have observations by both \fermi/GBM and \swift/BAT its luminosity varied over several orders of magnitude, making it an excellent test case for studying the evolution of the magnetospheric radius with accretion rate.
Upon the initial discovery of the system, the Gaia Data Release 2 estimated the distance of the NS as $6.8^{+1.5}_{-1.1}$~kpc \citep{2018AJ....156...58B}, that was adopted by most follow up studies \citep[e.g.][]{2018Natur.562..233V}. 
However, analysis of data from the Gaia Data Release 3 (DR3) yielded a distance of $5.2\pm0.3$~kpc \citep{2021AJ....161..147B}.
Finally, \citet{2020A&A...640A..35R} computed a distance of $5.5\pm1.7$~kpc based on BVRI photometric measurements to estimate the interstellar absorption. Assumptions about the distance on the source play an important role in the modelling of the NS spin evolution as we will see in the next sections.

\begin{table*}
\caption{Results of modelling \lxps and \rxjs.}
\begin{threeparttable}
\resizebox{\textwidth}{!}{%
\begin{tabular*}{\textwidth}[t]{p{0.05\textwidth}p{0.1\textwidth}p{0.0\textwidth}p{0.1\textwidth}p{0.0\textwidth}p{0.1\textwidth}p{0.0\textwidth}p{0.1\textwidth}p{0.0\textwidth}p{0.14\textwidth}p{0.0\textwidth}p{0.07\textwidth}p{0.05\textwidth}}
\hline\noalign{\smallskip}
Params\tnote{$\dagger$} & RX\,J0520 &  & RX\,J0520 & & RX\,J0520 & & RX\,J0209  &  &  RX\,J0209 & & RX\,J0209 &units\\
  &   (GL79) &  & (W95) & & Literature Values$^{1}$ & & (GL79) &  &   (W95) && Literature Values$^{2}$& \\
\hline\hline\noalign{\smallskip}
\multicolumn{4}{l}{Keplerian Orbit Parameters}\\\noalign{\smallskip}
		$e$ &  0.037$\pm$0.015& & 0.036$\pm$0.017 & & 0.029$\pm$0.010 & &0.324$\pm$0.016 &  & 0.321$\pm$ 0.011&& 0.319& --\\
		$P_{\rm orb}$  & 23.97$\pm$0.06& & 23.98$\pm$0.07 & & 23.93$\pm$0.07 & &47.16$\pm$0.21  &  & 47.39$\pm$ 0.17&& 47.37& d\\
		$\omega$ & 229$\pm$31&  &   226$\pm$38& & 233$\pm$18 & &79.5$\pm$3.3&  & 77.1$\pm$ 2.4&& 65.7& $^o$\\
		$a \sin i$ &  105.1$\pm$1.6 & &  105.0$\pm$1.8& & 107.6$\pm$0.8 & &162$\pm$3 &  &  164.1$\pm$2.1&& 169.8 & lsec\\
		$T_{\rm \pi/2}$  &  56666.40$\pm$0.07&  & 56666.89$\pm$0.08 & & 56666.41$\pm$0.03 & & 58793.7$\pm$0.4 & & 58792.84$\pm$0.23&& 58785.76 & MJD \\
\hline\noalign{\smallskip}
\multicolumn{4}{l}{Torque model Parameters}\\\noalign{\smallskip}
        $\log{B}$  & 11.688$\pm$0.011&  &  11.878$\pm$0.010 & & -- & &11.875$\pm$0.003  &  & 12.0767$\pm$0.0022&& -- & G\\
        $v_{0}$ & 124.3921$\pm$0.0008$^{3}$&  &  124.3920$\pm$0.0009$^{3}$& & -- & &107.4911$\pm$0.0007$^{4}$  &  &107.4913 $\pm$0.0005$^{4}$&& --& mHz\\
		$\xi_{0}$ & 0.5$^*$&  &  0.5$^*$&  &-- & & 0.5$^{*}$& &0.5$^{*}$&& -- & --\\
\hline\noalign{\smallskip}
\multicolumn{4}{l}{Other Parameters}\\\noalign{\smallskip}
    $d$  & 50\tnote{$\ddagger$}&  & -- &  & -- &  & 62\tnote{$\ddagger$} & & -- && -- & kpc\\
	$\ln{f}$ & -13.10$\pm$0.17&  &  -13.00$\pm$0.17 & & -- & &-13.10$\pm$0.13& & -13.41$\pm$0.13&& -- & --\\
\hline\noalign{\smallskip}
\multicolumn{4}{l}{Evidence}\\\noalign{\smallskip}
	$\ln{Z}$ & 302.9$\pm$0.4 &  &  300.3$\pm$0.6 & & -- & &492.3$\pm$0.7& &504.0$\pm$0.6&& -- & --\\
\hline\noalign{\smallskip}
\end{tabular*}
} 
\footnotesize
$\dagger$ Reported values of the fitted parameters and their uncertainties are estimated from the mean and standard deviation of the constructed posterior samples.
$\ddagger$ Distance was fixed to the LMC/SMC values.
$^{1}$
\citet{2020ApJ...896...90M}
$^{2}$ GBM Accreting Pulsars project: \url{https://gammaray.msfc.nasa.gov/gbm/science/pulsars.html}
$^{3}$ Reference MJD: 56645.3
$^{4}$ Reference MJD: 58807.0
\end{threeparttable}
\label{tab:RXJ_results}
\end{table*}

\section{Results}\label{sec:results}
We present the results of our analysis for the three systems introduced in the previous section.

\subsection{RX\,J0520.5-6932 (LXP 8.04)}
\begin{figure}
  \includegraphics[width=\columnwidth]{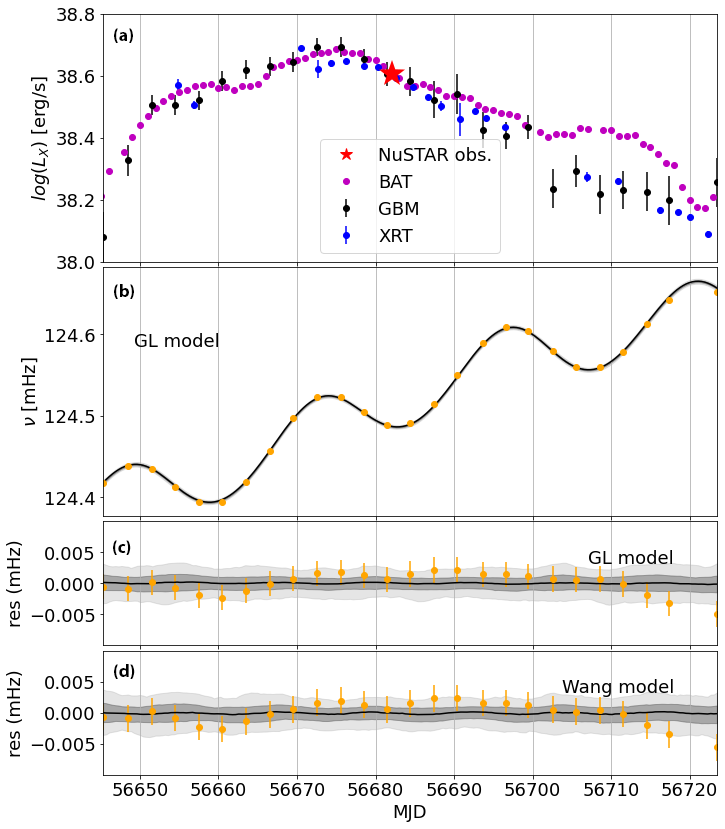}
    \caption{Panel (a): Temporal evolution of the X-ray luminosity of \lxps as measured by the \swift/BAT and \fermi/GBM all-sky surveys,  
    and \swift/XRT monitoring. The luminosity derived by a single \nustar observation is also marked with a star. Panel (b): Spin frequency evolution as measured by \fermi/GBM and
    fitting results
    using the \protect\citetalias{1979ApJ...234..296G} model. 
    Panel: (c) Residuals of the fit using the \protect\citetalias{1979ApJ...234..296G} torque model. 
    Panel: (d) Same as panel (c) but for the \protect\citetalias{1995ApJ...449L.153W} torque model. The dark and light grey shaded regions indicate respectively the $68\%$ and $99.5\%$  ranges of our solutions. Uncertainties in points that appear in the residual plots are based on \fermi/GBM measurements.}
    \label{fig:RXJ0520}
\end{figure}

To estimate the mass accretion rate as a function of time we used the standard methods described in Sec. \ref{sec:obsdata}. We first obtained the XRT and BAT count rates and pulsed flux from GBM. Then we used the bolometric X-ray luminosity obtained by \nustar data \citep[i.e. $L_{\rm X} =$ 4\ergs{38} at MJD 56682,][]{2014ApJ...795..154T}  to scale GBM data. This resulted in the light curves shown in the upper panel of Fig. \ref{fig:RXJ0520}. The conversion factors we used are $6.55\times10^{40}\:{\rm erg\:s^{-1}\:count^{-1}\:s}$, $3.2\times10^{39}\:{\rm erg\:s^{-1}\:count^{-1}\:s}$ and $4.2\times10^{37}\:{\rm erg\:s^{-1}\:count^{-1}\:s}$ for BAT count rates, GBM pulsed fractions and XRT count rates respectively. The inferred light curves from all three instruments agree for the bright luminosity state. However, when the luminosity drops below about 2.5\ergs{38}, estimates based on GBM overshoot both XRT and BAT measurements. Another interesting feature is that BAT and XRT estimates are in good agreement between them when data from both instruments are available.
Given that GBM pulsed fractions can be affected by changes in the pulse profile we opted to use the 
BAT data as proxy for the bolometric luminosity and the inferred mass accretion rate. 

Having an estimate for the mass accretion we applied our recipe and fitted the data using the \protect\citetalias{1995ApJ...449L.153W} and \protect\citetalias{1979ApJ...234..296G} models. As an example, we show in Fig.~\ref{fig:RXJ0520}~(b) the fitting result to the measured spin frequencies (including modulation because of the orbital motion) using the \protect\citetalias{1979ApJ...234..296G} model. 
The evidence of the \protect\citetalias{1995ApJ...449L.153W} model is $\ln{Z}=$300.3 versus 302.9 for the \protect\citetalias{1979ApJ...234..296G} (see Table \ref{tab:RXJ_results}), which translates to the latter being $\sim10$ times more probable than the \protect\citetalias{1995ApJ...449L.153W} model, assuming the models are equally probable a priori.
The posterior distributions of the orbital solution and the \protect\citetalias{1979ApJ...234..296G} model parameters are presented in Fig.~\ref{fig:RXJ0520corner}.
The orbital parameters we recovered by both models are close to the estimates from previous works related to this system (see Table \ref{tab:RXJ_results}). Our orbital solution is also in agreement with the one presented in \citet{2020ApJ...896...90M} where the GBM pulse profiles phase offsets were modelled to refine the orbital solution of the source. Most importantly our method enables estimation of each model parameter and their uncertainties more accurately than the standard least-square minimization method \citep[e.g.][]{2017PASJ...69..100S}.

Based on the \protect\citetalias{1995ApJ...449L.153W} model we estimated a polar magnetic field of $\sim1.6\times10^{12}$ G for the NS. This is in agreement with other direct measurements of $B$.
In particular, the study of the broadband spectrum of \lxps by \nustar also revealed the presence of a  cyclotron resonance scattering feature 
at $\sim31.5$ keV yielding a direct measurement of $B\sim2\times10^{12}$~G \citep{2014ApJ...795..154T}.

\begin{figure}
 \includegraphics[width=\columnwidth]{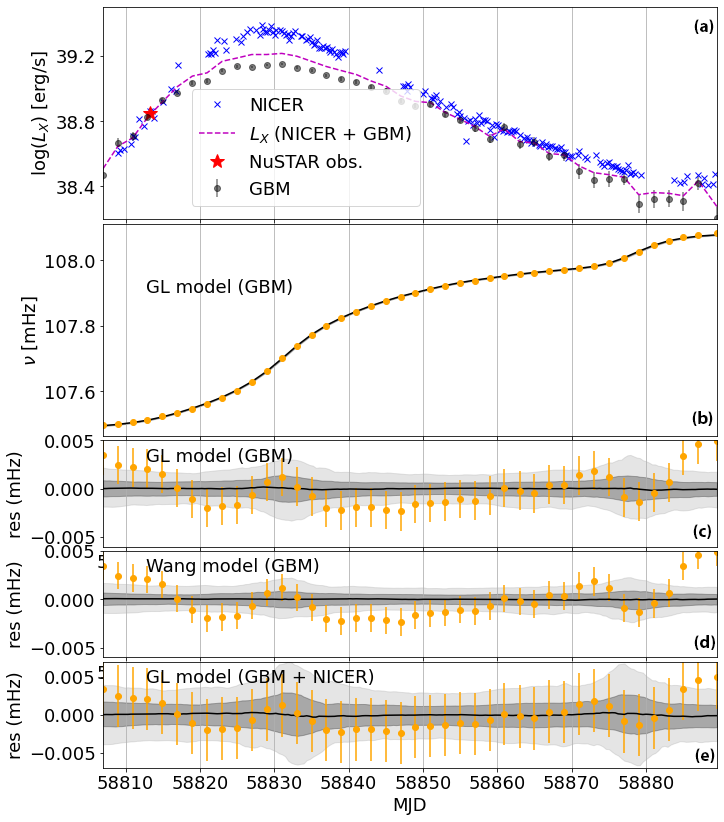}
    \caption{Panel (a): 
    Temporal evolution of the luminosity of \rxjs as measured by the \fermi/GBM all-sky survey and the \nicer observatory. Panel (b): 
    Results of fit to the observational data using the \protect\citetalias{1979ApJ...234..296G} model and \fermi/GBM pulsed fractions as a proxy for the luminosity. Panel (c): 
    Residuals of the fit using the \protect\citetalias{1979ApJ...234..296G} model and \fermi/GBM pulsed fractions as a proxy for the luminosity. Panel (d): Same as in panel (c) but for     \protect\citetalias{1995ApJ...449L.153W} model.
    Panel (e): Same as in panel (c) 
    but for a combination of \fermi/GBM pulsed fractions and \nicer measurements as a proxy for the luminosity.}
    \label{fig:RXJ0209}
\end{figure}
\subsection{RX\,J0209.6-7427}
To estimate the mass accretion rate as a function of time we used the methods introduced in Sec. \ref{sec:method}. First, we obtained \nicer luminosity measurements and pulsed fluxes from GBM which we also used as a proxy for the $L_{\rm X}$. We modelled the luminosity using two methods, first using both the \nicer and GBM data and then using only the GBM pulsed flux. As evident by comparing panels (c) and (d) of Fig.~\ref{fig:RXJ0209}  the latter method yields better results.
In fact, comparing the bolometric luminosities scaled 
using the GBM or \nicer energy ranges alone (see Fig.~\ref{fig:RXJ0209} upper panel) we see that the \nicer $L_{\rm X}$ is systematically higher for the brightest phase of the outburst. 
This could be a result of a contribution to the \nicer band from disc soft X-ray radiation, thus leading to an overestimation of the peak value of $L_{\rm X}$. 
A similar excess due to contribution from a soft component has also been reported in other super-Eddington accreting systems and has been proposed to be related to the hot accretion disc and/or outflows \citep{2019ApJ...873...19T,2020MNRAS.491.1857D}.

Comparing the \protect\citetalias{1979ApJ...234..296G} and \protect\citetalias{1995ApJ...449L.153W} models, we find that the latter yields better results, with $\ln{Z} = 504.0$ as compared to $492.3$ for the \protect\citetalias{1979ApJ...234..296G} model (see Table \ref{tab:RXJ_results}). Thus the \protect\citetalias{1995ApJ...449L.153W} model is $\sim2\times10^{5}$ times more probable than the \protect\citetalias{1979ApJ...234..296G} model, assuming the models are equally probable a priori. 
The model parameters are listed in Table \ref{tab:RXJ_results}, while in Fig. \ref{fig:RXJ0209_corner_Wang} we show the corner plot of posterior distributions for the better model.
The orbital parameters we recovered with the two models are similar to each other and to the estimates from previous works related to this system, with the exception of the semi-projected binary separation ($a\sin{i}$), which is found to be smaller than the previously published values.  

\subsection{\swj}\label{sec:SWIFTresults}
\begin{figure}
 \includegraphics[width=\columnwidth]{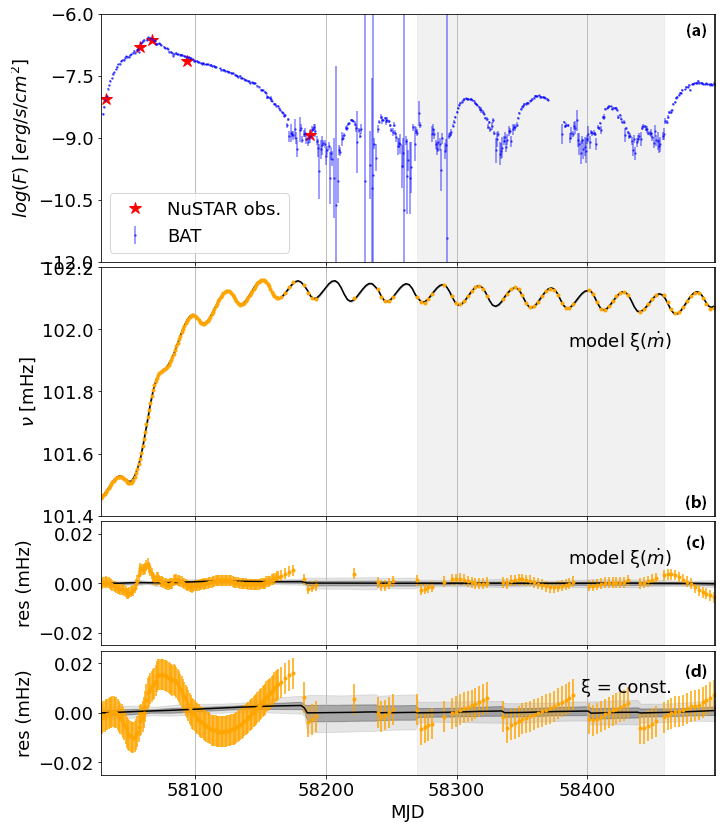}
 \vspace{-0.3cm}
    \caption{Panel (a): Temporal evolution of the flux of \swjs as obtained by scaling \swift/BAT count rates to the \nustar fluxes. For clarity, the bolometric flux is plotted instead of the luminosity, since the distance to the system is a free parameter. Panel (b): 
    Fitting results to the observational data (after removing the orbital effects) using the \citetalias{2014MNRAS.437.3664H} torque model and the mass-accretion dependent $\xi$ parameter of Eq.~(\ref{xi}). The orbital effects were removed by independently modelling the data obtained within the gray shaded region as described in the text.
    Panel (c):
    Residuals of the fit shown in panel (b). 
    Panel (d): Residuals of the fit using the standard $\xi =$const. approach and the \citetalias{2014MNRAS.437.3664H} torque model.}
    \label{fig:SWIFTJ0243}
\end{figure}

For the estimation of the system's bolometric X-ray flux, $F_{\rm X}$, we considered a linear relation with the  \swift/BAT count-rates ${\rm CR}_{\rm BAT}$, i.e. $F_{\rm X} = A\cdot {\rm CR}_{\rm BAT} $, where $A$ is determined as follows. \citet{2019ApJ...873...19T} calculated the flux at five dates (see panel (a) of Fig. \ref{fig:SWIFTJ0243}) 
using \nustar observations (see Table \ref{Tao}). Assuming that these are a good proxy of the bolometric flux, we performed a linear fit to those fluxes and the \swift/BAT count rates on the same days.  The slope\footnote{To estimate bolometric $L_{\rm X}$ we simply scale by the source distance $4\pi d^{2}$} of the linear fit was found to be $1.47\pm0.13\times10^{-7}$~erg cm$^{-2}$ s$^{-1}$ $(\rm counts/s)^{-1}$.
Contrary to the other systems we studied, which lie in the Magellanic Clouds, the distance to \swjs is more uncertain despite the very accurate parallax measurements by Gaia (see Sec. \ref{sec:application} for more details). Therefore, we treated $d$ as a free parameter,  allowing it to take values between 4 kpc and 8 kpc. For the estimation of the mass accretion rate from the $L_{\rm X}$ we used the same method as in the previous systems (for more details, see Sec. \ref{sec:obsdata}).

\begin{table}
	\centering
	\small
    \setlength\tabcolsep{2pt}
	\caption{X-ray fluxes of \swjs from \citet{2019ApJ...873...19T} with the corresponding dates and \nustar Observational Identification Number (ObsID).}
	\label{tab:Fluxes}
	\begin{tabular}{lccr} 
		\hline
		ObsID & $F_{3-79\, \rm keV}\:\rm (erg\:cm^{-2}\:s^{-1})$ & MJD\\
		\hline
		90302319002 & $8.73\times10^{-9} $ & 58031.661\\
		90302319004 & $1.55\times10^{-7}$ & 58057.306\\
		90302319006 & $2.27\times10^{-7}$ & 58067.105\\
		90302319008 & $0.74\times10^{-7}$ & 58093.615\\
		90401308002 & $1.14\times10^{-9}$ & 58187.515\\
		\hline
	\end{tabular}
    \label{Tao}
\end{table}

Initially, we tried to fit our complete data set (MJD 58027.5-58497.5) following a similar procedure as for \rxjs and \lxps. This approach was hampered mainly by two issues: (i) the appearance of gaps in the GBM frequency monitoring due to the source entering a faint state; (ii) in the low flux states the source intensity is not properly characterised by \swift/BAT or other all-sky monitoring programs. The typical methodology used is to assume a steady spin-down term during these epochs \citep[e.g.][]{2017PASJ...69..100S}. However, given that this spin-down term is a function of accretion rate, we adopted a brute-force approach where we introduce a ``jump'' in frequency for every large gap in the available data. In the \fermi/GBM data we identified 6 large gaps. We therefore added 6 extra free parameters in our model. 
The first results obtained by our MCMC approach (not shown here) revealed that a model with constant $\xi$ for the whole duration of the outburst yields large residuals and cannot explain the data. Moreover, due to the extra free parameters, it required excessive computational time to run tests with different torque models and expressions of $\xi(\dot{m})$. For these reasons, we opted to model the orbital modulation and intrinsic torque separately.

\begin{table}
	\centering
	\small
    \setlength\tabcolsep{2pt}
	\caption{Orbital parameters of \swjs.}
	\label{tab:Orbitals}
	\begin{tabular}{lccr} 
		\hline
		Params & Result & Literature Value$\dagger$ & Units\\
		\hline
		$e$ & 0.0987 $\pm$ 0.019 & 0.103 &-\\
		$P_{\rm orb}$ & 27.693 $\pm$ 0.005  &27.70 &d\\
		$\omega$ & -74.9 $\pm$ 1.1 &-74 &$^{o}$\\
		$a \sin i$ & 116.43 $\pm$ 0.22 & 115.5& 1-sec\\
		$T_{\rm \pi/2}$ & 58115.63 $\pm$ 0.04 & 58115.6& MJD\\
		\hline
	\end{tabular}
	\begin{tablenotes}
      \small
      \item$\dagger$ GBM Accreting Pulsars project.
    \end{tablenotes}
    \label{SWIFTJorbitals}
\end{table}

\begin{figure}
 \includegraphics[width=\columnwidth,clip]{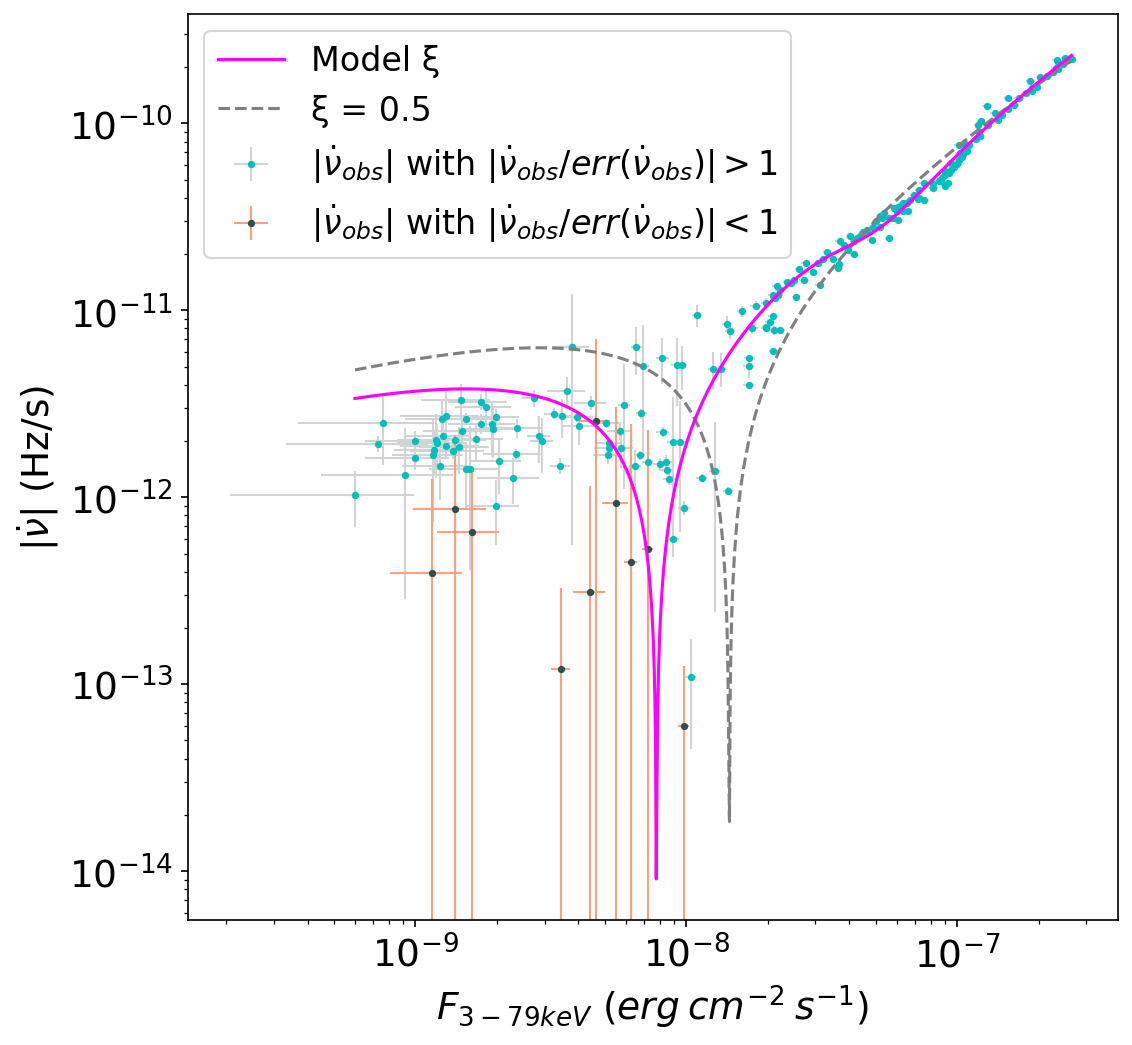}
 \vspace{-0.5cm}
     \caption{Plot of the absolute value of the frequency derivative $|\dot{\nu}|$ versus the X-ray flux $F_{\rm 3-79keV}$ for \swjs. Symbols indicate the observational values from \fermi/GBM, the solid magenta line shows the prediction of our best-fit model with $\xi(\dot{m})$ given by Eq.~(\ref{xi}) and the dashed line shows the best-fit model using the standard approach with a constant $\xi=0.5$. }
    \label{fig:SWIFTJ0243-vdot}
\end{figure}

First, to derive the orbital parameters we used a chunk of observational data (MJD 58260 to 58460), where no large fluctuations of the flux were apparent, to calculate and remove the orbital effects from our problem, considering a constant $\xi \approx 0.5$. The orbital parameters we recovered fall within $1\sigma$ from the results obtained from previous works on that system (see Table \ref{SWIFTJorbitals}). Having removed the orbital modulation from our data, we proceeded with the modelling of the intrinsic NS spin up. Instead of estimating the spin-up rate using a torque model and then fitting it to GBM observed frequencies (as we did in the previous systems), we calculated the derivative of the GBM frequency data and fitted the theoretical spin-up rate predicted by our model to the $F_{\rm X}-\dot{\nu}$ space (we remind that distance is a free parameter). This method is more efficient, because multiple intermediate steps from our process can be eliminated from every run of our algorithm. Nevertheless, the results of the fit performed to $\nu(t)$ (similar to \lxps and \rxjs) or to the $F_{\rm X}-\dot{\nu}$ space are consistent to each other.

We fitted the $F_{\rm X}-\dot{\nu}$ data using the \citetalias{2014MNRAS.437.3664H} torque model, since the other models cannot describe both the low and high $L_{\rm X}$ regimes (for more details, see Sec.~\ref{sec:models}). We then considered two cases, one with $\xi=0.5$ and another one with a physically motivated parametrization of $\xi$ on the accretion rate, as described in Sec.~\ref{sec:method} (see Eq.~\ref{xi}). The corresponding results are presented in Table \ref{tab:SWIFTresults-2} and in panels (b) to (d) of  Fig.~\ref{fig:SWIFTJ0243}, where  
we compare the \fermi/GBM frequencies with the best-fit model and the residuals of both models for $\xi$. Inspection of panels (c) and (d) shows that a model with constant $\xi$ yields larger residuals than the parametric $\xi$ model of Eq.~(\ref{xi}). The $\ln{Z}$ factor of the latter model was estimated to be 5506.5 as compared to 5417.7 for a constant $\xi$ (see Table \ref{tab:SWIFTresults-2}), meaning that the  $\xi(\dot{m})$ model is $\sim 10^{38}$ times more probable, assuming the both are equally probable a priori. 
In our model the magnetic field is calculated indirectly from the $a_{\rm 0}$ parameter, as described in Eq.~(\ref{Bmodel}). The result from the fit was $\log (B/{\rm G}) = 13.061 \pm 0.017$, or a polar magnetic field of $\sim2.3\times10^{13}$ G.
Fig.~\ref{fig:SWIFTJ0243-vdot} shows the best fit-solution for the constant $\xi$ and $\xi(\dot{m})$ models in the $F_{\rm X}-\dot{\nu}$ space, with the $\dot{\nu}$ measurements of  \fermi/GBM overplotted for comparison (see Fig. \ref{fig:SWIFTJ_2_corner} for posterior distributions). We note that the variable $\xi$ model can explain quite well the transition between the high and low luminosity regimes\footnote{During the revision of this manuscript we became aware of an independent study \citep[i.e.][]{2022MNRAS.512.5686L} that also noted a so-called flattening in the $\dot{\nu} - L_{\rm X}$ space of \swjs. Same flattening effect would be evident in out Fig. \ref{fig:SWIFTJ0243-vdot} if plotted in linear scale. We note that our results quantitatively match their findings.}.

At this point we should comment that the updated distance by Gaia (i.e. Gaia DR3, $d=5.2\pm0.3$ kpc) does not fall within the 3$\sigma$ range of values derived by torque modelling (see Table \ref{tab:SWIFTresults-2}). In fact, if we fix the distance at 5 or 6 kpc the data cannot be fitted by our model using standard NS parameters (see further discussion in Sec.~\ref{sec:Discussion-WhatDidWeLearn}). However, if we treat the NS mass, radius and moment of inertia as free parameters and set a hard limit on the distance at 6 kpc we are able to get acceptable fits to the data (see Table \ref{tab:SWIFTresults-2} and Figure \ref{fig:SWIFTJ_MIR_corner}).

\begin{table*}
\caption{Results of modelling \swj in the $F_{\rm X} - \dot{\nu}$ space.}\begin{threeparttable}
\begin{tabular*}{\textwidth}[t]{p{0.2\textwidth}p{0.2\textwidth}p{0.2\textwidth}p{0.2\textwidth}p{0.20\textwidth}}
\hline\noalign{\smallskip} 
Parameters\tnote{$\dagger$} &$\xi =$const. &   $\xi(\dot{m})$ model & $\xi(\dot{m})$ model, with &units\\
                        &              &                        & $M,\,R\,$and $I\,$ as free parameters&      \\
\hline\hline\noalign{\smallskip} 
\multicolumn{5}{l}{Torque model Parameters}\\\noalign{\smallskip} 	
        $\log{B}$  & 13.349$\pm$0.013&   13.061 $\pm$ 0.017\tnote{*} & 12.793$\pm$ 0.029\tnote{*}& G\\
		$\xi$ &  0.50&     -- & -- &--\\
		$a_{\rm 0}$ & -- &      3.608$\pm$0.010& 3.612 $\pm$ 0.008 &--\\
		$a_{\rm 1}$ & -- &     0.090$\pm$0.009 & 0.080 $\pm$ 0.006 &--\\
		$a_{\rm 2}$ & -- &     -0.20$\pm$0.03 & -0.05 $\pm$ 0.03 & --\\
		$a_{\rm 3}$ & -- &     4.4$\pm$0.9 & 5.6 $\pm$ 1.2 &--\\
\hline\noalign{\smallskip} 	
\multicolumn{5}{l}{NS Parameters}\\\noalign{\smallskip}
    $M$ & 1.4 & 1.4 & 1.116 $\pm$ 0.015 &$M_{\sun}$\\
	$R$ & 1.2 & 1.2 & 1.294 $\pm$ 0.006 &$10^{6}\,{\rm cm}$\\
	$I$ & 1.3 & 1.3 & 1.005 $\pm$ 0.004 &$10^{45}\,{\rm g\,cm^{2}}$\\
\hline\noalign{\smallskip}
\multicolumn{5}{l}{Other Parameters}\\\noalign{\smallskip}
    $d$  & 6.99 $\pm$ 0.03  &   7.47$\pm$0.06 & 5.986 $\pm$ 0.013 & kpc\\
	$\ln{f}$ & -25.65$\pm$0.05&    -26.10 $\pm$ 0.05& -26.09 $\pm$ 0.05 & --\\
\hline\noalign{\smallskip}
\multicolumn{5}{l}{Evidence}\\\noalign{\smallskip}
	$\ln{Z}$ & 5419.27$\pm$0.25 &   5507.6 $\pm$ 0.3& 5495.0 $\pm$ 0.5 & --\\
\hline\noalign{\smallskip}  
\end{tabular*}
\begin{tablenotes}\footnotesize
\item[$\dagger$] Reported values of the fitted parameters and their uncertainties are estimated from the mean and standard deviation of the constructed posterior samples.
\item[*] Inferred from $a_{\rm 0}$, $M$ and $R$.
\end{tablenotes}
\end{threeparttable}\label{tab:SWIFTresults-2}
\end{table*}

\section{Discussion}
\label{sec:dis}
\subsection{Application to systems in the Magellanic Clouds}

We have studied the properties of two BeXRB systems  in the Magellanic Clouds using different torque models \citep[i.e.][]{1979ApJ...234..296G, 1995ApJ...449L.153W}, based on X-ray data collected during their outbursts. We used data from \nicer, \fermi/GBM, XRT and \swift/BAT as a proxy for the luminosity and then scaled our time-series using \nustar observations to retrieve the bolometric $L_{\rm X}$. With that data we were able to get an expression for the theoretically predicted spin evolution of our sources, and to obtain posterior distributions for the magnetic field and the orbital parameters of our systems, using {\sc ultranest} to fit to the \fermi/GBM frequency data. 

\begin{figure}
 \includegraphics[width=\columnwidth]{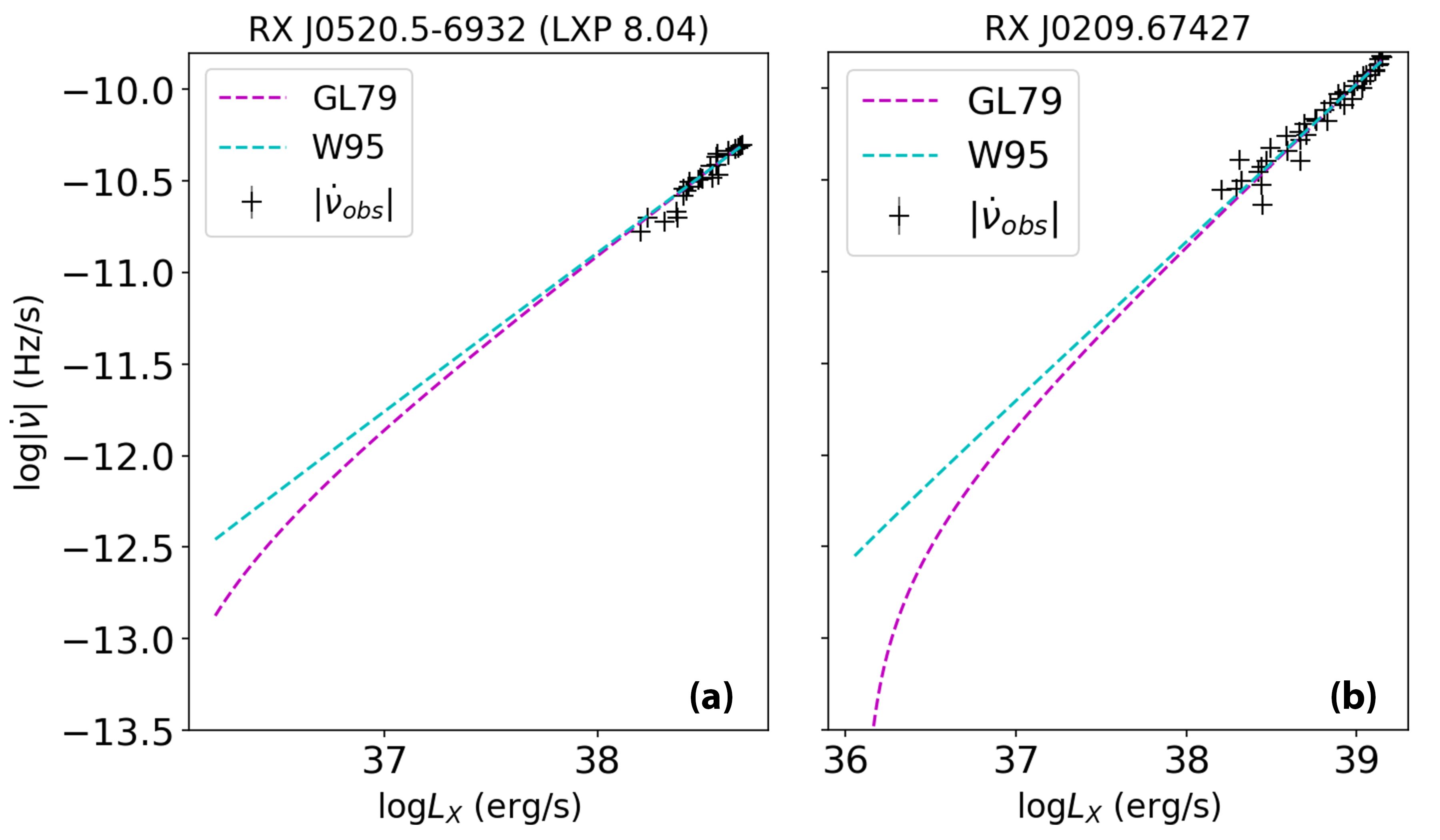}
 \vspace{-0.5cm}
    \caption{Panel (a): Plot of the absolute value of the frequency derivative $|\dot{\nu}|$ versus the bolometric X-ray luminosity $L_{\rm X}$ for \lxps. Symbols indicate the observational values from \fermi/GBM, while the dashed lines show the prediction of the \protect\citetalias{1979ApJ...234..296G} and \protect\citetalias{1995ApJ...449L.153W} models. Panel (b): Similar to (a), but for \rxjs. }
    \label{fig:RXJ-dotv-Lx}
\end{figure}

For \lxps and \rxjs we were able to simultaneously obtain orbital solutions for the binary and estimates on the $B$ field of the NS. 
Most importantly, by using the full expression of the torque models, we are not limited to the asymptotic behaviour during $\omega_{\rm fast} \ll 1$, which is commonly used in the literature \citep[e.g.][]{2017PASJ...69..100S,2017ApJ...843...69W}, but we also explore behaviours where the slope changes in a $\log{|\dot{\nu}|}$ versus $\log{L}_{\rm X}$ diagram at lower accretion rates as we approach equilibrium.
This approach enables to test which of the \protect\citetalias{1979ApJ...234..296G} and \protect\citetalias{1995ApJ...449L.153W} torque models can better explain the data. Thus we can favor one model over the other for \rxjs and somewhat less prominently in the case of \lxps (see also Table~\ref{tab:RXJ_results}). 
In Fig. \ref{fig:RXJ-dotv-Lx} we plot $|\dot{\nu}|$ versus $\dot{M}$ together with both models with their best fit parameters. 
However, the question about which torque model is favorable over the other is not entirely solved. There are other systematic uncertainties, like a luminosity dependent bolometric correction factor \citep[e.g.][]{2022MNRAS.513.1400A}, which could reduce the amount (i.e. $ln(Z)$) one model is favored over another.

The problem of accretion disc threading by stellar magnetic field still lacks a comprehensive solution as demonstrated by theoretical and observational studies \citep[e.g.][]{2009A&A...493..809B,2017AstL...43..706F,2020ApJ...896...90M}.
Nevertheless, regardless of the torque model, it is possible to relax some of the underlying assumptions or introduce extra terms to create wider or narrower cusp-like transitions around equilibrium. For example by assuming a misalignment between magnetic and rotation axes it is possible to induce a sharper transition near equilibrium \citep[see][for an application to 4U\,1626-67]{2020MNRAS.495.3531B}.

\subsubsection{Implications from MC depth estimations}

\begin{figure}
 \includegraphics[width=0.49\columnwidth]{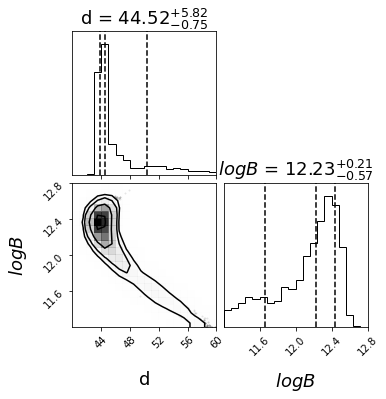}
 \includegraphics[width=0.49\columnwidth]{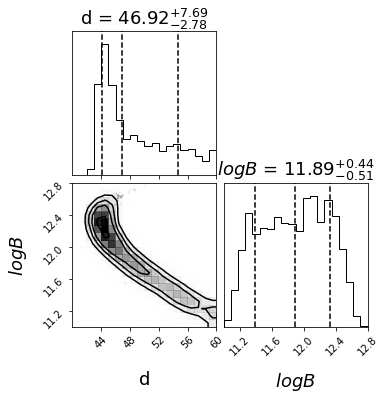}\\
 \includegraphics[width=0.49\columnwidth]{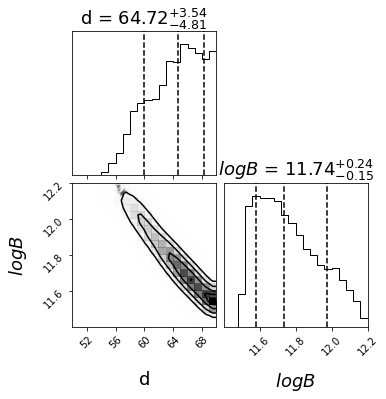}
 \includegraphics[width=0.49\columnwidth]{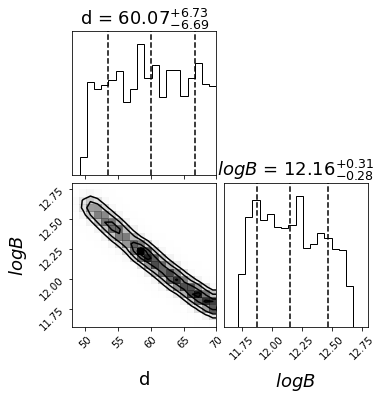}
    \caption{\emph{Top:} Contour plots is calculated by fitting the \protect\citetalias{1979ApJ...234..296G} (left panel) and \protect\citetalias{1995ApJ...449L.153W} (right panel) to the $\dot{M}-\dot{\nu}$ parameter space for RX\,J0520. For the fit we fixed orbital parameters and kept the magnetic field and distance $d$ as a free parameter (40-60 kpc prior). \emph{Bottom:} Same as top but for RX\,J0209.}
    \label{fig:RXJ-dist}
\end{figure}

In the calculations concerning sources in the Magellanic Clouds a possible source of uncertainty is distance.
While the average distances of the SMC and LMC are well determined, the depth of each galaxy is of the order of 4-8 kpc \citep{2009A&A...496..399S}. Thus, it is prudent to at least explore if the fit to the data sets of the two systems can improve by treating the distance to each source as a free parameter. For simplicity we fitted the model to the $\dot{M}-\dot{\nu}$ parameter space, although performing the fit on a similar manner as above yielded the same results and trends in the corner plots.
The results are shown in Fig. \ref{fig:RXJ-dist}. For \lxps we found that the data favour a somewhat smaller distance than the average one of LMC, which is however still consistent with the depth of the galaxy. 
Interestingly, for \rxjs regardless of the torque model used we find is a degeneracy between the distance and the magnetic field, meaning that we cannot put any constraints on the position of the system compared to the average distance of the nearby galaxies.
The application to \rxjs also demonstrates how the uncertainty in distance affects the $B$-field estimates.
For example, using a uniform prior for the distance between 50-70 kpc, the 1 $\sigma$ uncertainty in the derived magnetic field strength is about 3 or dex of 0.5.

Recently, an independent study of the spin-up of \rxjs with the use of a generalised torque model (i.e. $\dot{\nu}\propto L^a$) revealed a somehow steep dependence (i.e. $a=1$) of spin-up rate on luminosity \citep{2022MNRAS.517.3354L}. Considering that the authors used a fixed distance of 55 kpc, this would be consistent with our results.

\subsection{What did we learn for the first Galactic PULX?}\label{sec:Discussion-WhatDidWeLearn}

\begin{figure}
 \includegraphics[width=\columnwidth]{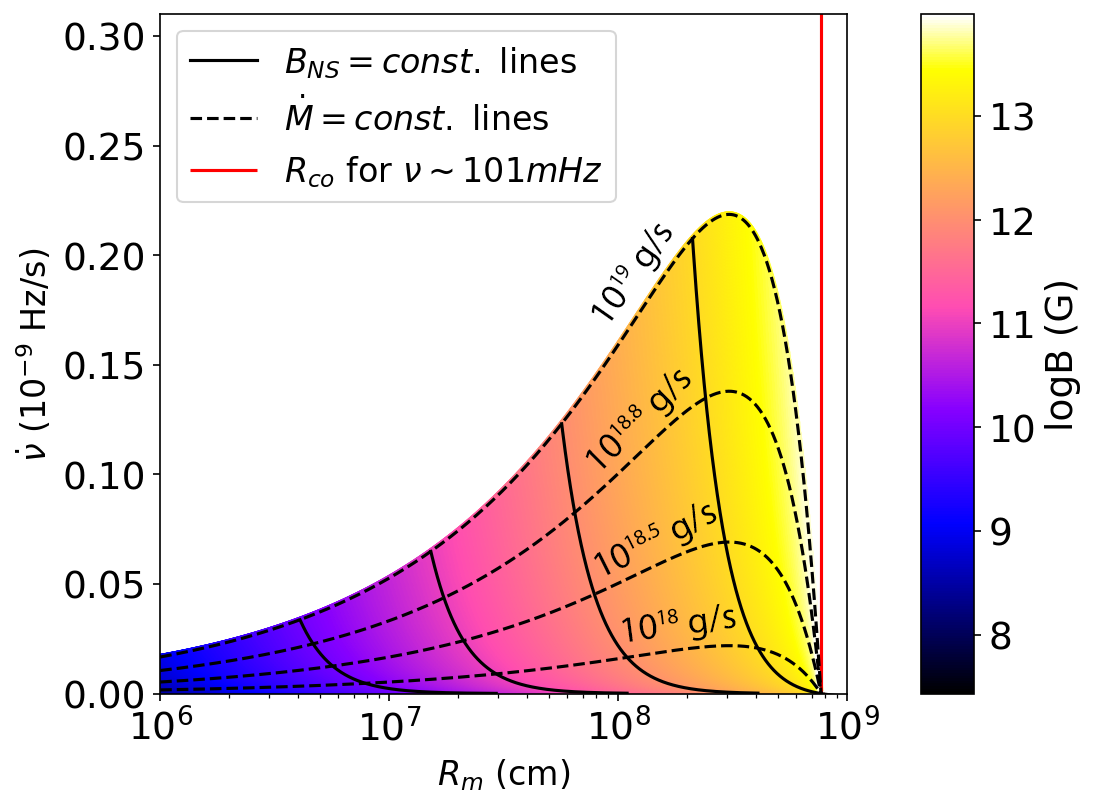}
 \vspace{-0.5cm}
    \caption{Induced spin-up rate assuming a disc truncated at $R_{\rm m}$ for various accretion rates (dashed lines).
    A fiducial NS during an outburst can only move along paths of constant $B$ (solid lines) for a given torque model (here the \citetalias{2014MNRAS.437.3664H} model was used). From left to right we have plotted the paths for $10^{10}$ G, $10^{11}$ G, $10^{12}$ G and $10^{13}$ G. The vertical solid red line indicates the corotation radius assuming a NS spin frequency of 101 mHz.
    }
    \label{fig:dotv-Rm}
\end{figure}

For \swjs we adapted the method used for the other two pulsars to tackle the complexity of the dataset. Given the dynamic range in the observed luminosity and the transitions between spin-up and spin-down phases we opted for using a more generalised form for the \citetalias{2014MNRAS.437.3664H} torque model.

One of the main difficulties arising when trying to create an empirical model for $\xi(\dot{m})$ is the lack of a way to attain direct measurements of the magnetospheric radius. As a result, we have to rely on comparing how well different methods describe our observational data, namely the luminosity and the NS spin. A way to overcome this difficulty is using a torque model and solving backwards for $R_{\rm m}$. This is possible using the \protect\citetalias{2014MNRAS.437.3664H} model, which can be successfully solved for $R_{\rm m}$ without making any assumptions for the form of the $\xi(\dot{m})$ or the magnetic field, provided that we have $\dot{v}$ and $\dot{m}$ measurements.

\begin{figure}
 \includegraphics[width=\columnwidth]{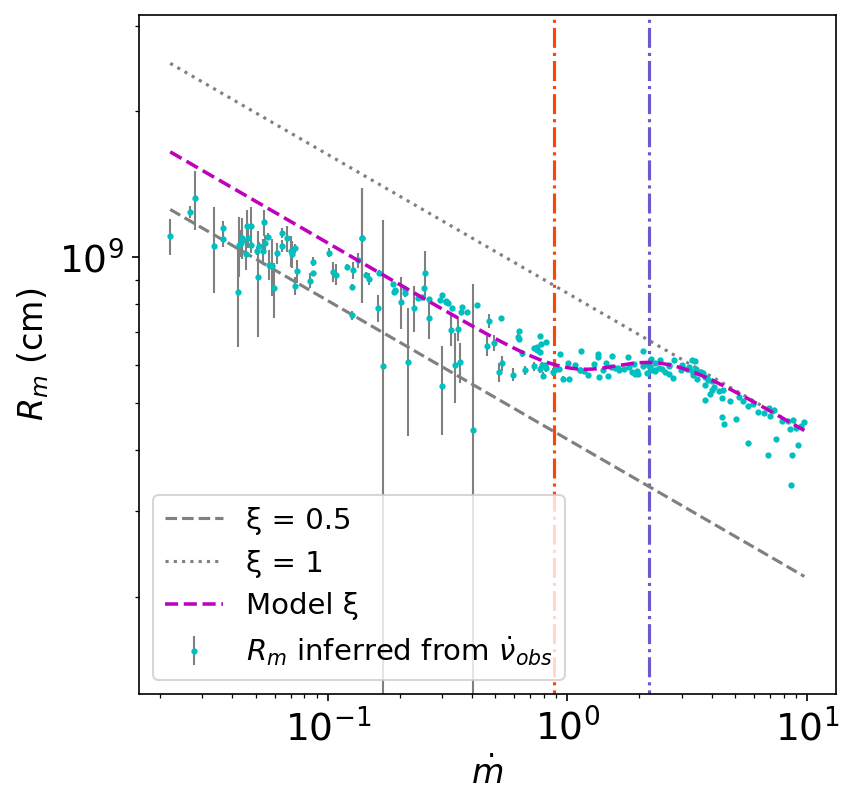}
     \caption{Plot of the magnetospheric radius versus the dimensionless accretion rate for \swjs. Symbols indicate the values inferred by solving Eq.~(\ref{eq:rm-direct}). The most probable model obtained by the Bayesian method is overplotted with a dashed magenta line. The vertical dash-dotted lines represent the change of the pulse profile of the pulsar from single peaked
     to double peaked and the transition from gas-pressure dominated to radiation-pressure dominated disc respectively from left to right (see Sec. \ref{sec:Discussion-WhatDidWeLearn}). 
     }
    \label{fig:SWIFTJ0243-Rm}
\end{figure}

We can create a set of observational $\dot{v}$ and $\dot{m}$ values for \swjs by calculating the gradient of the GBM  frequency data and assuming a distance of 6.95~kpc which is the result we got from our fit (see Table \ref{tab:SWIFTresults}) in order to be able to compare the data we generate to our model. Now, after interpolating the two sets of data to the same dates, using a linear interpolation method, we can solve the equation 
\begin{equation}
    \frac{1}{2\pi I}\:N_{\rm tot}(\dot{M}_{\rm obs},\:R_{\rm m}) - \dot{\nu}_{\rm obs} = 0
    \label{eq:rm-direct}
\end{equation}
for $R_{\rm m}$. This equation has two solutions for the spin-up phase, which correspond to the two intersection points of  a horizontal line (at a given value of $\dot{\nu}$) with a dashed curve shown in Fig. \ref{fig:dotv-Rm}). The non acceptable solutions in the spin-up phase can be easily identified after plotting them on a $R_{\rm m}-\dot{m}$ graph, since they fall away from the standard $R_{\rm m}(\dot{m})=\xi R_{\rm A}(\dot{m})$  (with $\xi = 0.5 - 1$) solutions by several orders of magnitude. 
Using this method we generated a set of $R_{\rm m}$ data points derived from the observational data, making no assumptions for the dependency of the magnetospheric radius on the accretion rate or the magnetic field other than the ones inherent in the torque model we used (\citetalias{2014MNRAS.437.3664H}). The results of this method are portrayed in Fig. \ref{fig:SWIFTJ0243-Rm}.

The most intriguing result for \swjs is that we found evidence of an evolving disc in qualitative agreement with theoretical predictions \citep[i.e.][]{2019A&A...626A..18C}. 
We can clearly see from our results that the magnetospheric radius is not to-scale with the Alfv\'en radius at high accretion rates, in contradiction with the standard torque models' predictions.
Our model describes very well the magnetospheric radius evolution at super-Eddington accretion rates (see Fig. \ref{fig:SWIFTJ0243-Rm}).
However, we should stress the degeneracy between the range of $\xi$ values and $B$ in our approach (see Eq. \ref{xi}).
This degeneracy is the same to the constant $\xi$ approximation that is evident if we leave both $\xi$ and $B$ free parameters (see Fig. \ref{fig:SWIFTJ0243_corner_const_xi}).
This introduces extra uncertainty in our estimation of $B$ apart from the statistical uncertainty derived from the fit.
For our estimations we opted for setting $\xi_{\rm max} = 1$ based on the upper limits on $\xi$ reported in the literature. 
We could instead set a lower limit on $\xi$ used for standard disc accretion, i.e. $\xi_{\min}=0.5$.  
Fixing the lower $\xi$ value to 0.5 yields $\log{B} {\rm (G)} = 13.43 \pm 0.08$. Thus, the range of the accepted $B$ values is $(2-5)\times10^{13}$~G. This estimate value is in agreement with the detection of a cyclotron resonance scattering feature between 120-146 keV in the \textit{insight-HMXT} spectra \citep{2022ApJ...933L...3K}.

Since we argue that our findings for an evolved $\xi$ may be a result of changes in the disc, it is interesting to compare the transitions found here with independent studies. In particular, a sharp state transition in the spectral and temporal properties of the system has been reported based on \textit{insight-HMXT} observations \citep{2020MNRAS.491.1857D}. Based on the study of power-spectra and quasi periodic oscillations (QPOs) a transition in the pulse profile was found to occur at about two times the Eddington limit. 
This transition was proposed to mark the border between gas-pressure dominated
and radiation-pressure dominated regions of the disc. 
Another interesting transition marks the change of the pulse profile of the pulsar from single peaked to double peaked \citep[see][]{2018ApJ...863....9W,2020MNRAS.491.1857D}. This critical transition has been attributed to the formation of the accretion column \citep[see][]{2012A&A...544A.123B} and has been used as a proxy for an indirect estimate of the magnetic field. For \swjs this transition was found at $\dot{m}\sim 1$ or $L_{\rm X}\sim L_{\rm Edd}$ for the distance of 7~kpc that we computed from our model \citep[see][]{2018ApJ...863....9W,2020MNRAS.491.1857D}.
For comparison purposes with mark these transitions with vertical lines in Fig.~\ref{fig:SWIFTJ0243-Rm}.

As we mentioned earlier the updated Gaia distance of \swjs introduces difficulties in finding a torque model that can fit the observed \fermi/GBM data (see Sec.~\ref{sec:SWIFTresults}).  
This is because a smaller distance yields a lower maximum $L_{\rm X}$ and lower mass accretion rate estimates. Thus, to explain the highest observed spin-up rates at the peak of the outburst a larger magnetic field strength is required (see Fig. \ref{fig:dotv-Rm}). This increases further the magnetospheric radius and, in our case, pushes it very close to the corotation radius, prohibiting essentially any further spin up. A way around this problem was to let the NS parameters free. Indeed, a good fit was found for a NS with larger radius and smaller mass than the typically assumed values (see Table \ref{tab:SWIFTresults-2} and Fig. \ref{fig:SWIFTJ_MIR_corner}). 
Searching the literature for NSs with reliable mass estimates \citep{2016ARA&A..54..401O}, the double NS system J0453+1559 hosts the NS with the smallest measured mass of 1.174(4) $M_{\odot}$ \citep{2015ApJ...812..143M}.
Moreover, a recent study of the isolated NS in the center of supernova remnant HESS J1731-347 \citep{2022NatAs.tmp..224D} indicated that the NS may be extremely light with having a mass of $0.77^{+0.20}_{-0.16}$~M$_{\odot}$  (1 $\sigma$ errors).
Thus, \swjs could potentially host a very low mass NS.
However, this approach yielded a magnetic field strength smaller (by a factor of 2) than the one inferred by the reported cyclotron line \citep{2022ApJ...933L...3K}. This would potentially mean that the cyclotron line is formed in regions with multi-polar magnetic field components \citep[e.g. see evidence of such configuration][]{2019ApJ...887L..21R,2020ApJ...893L..38C}, compared to the torques that are associated to the dipole component. 
Alternatively, one needs to revise the assumptions of our model and in particular the radiative efficiency of the accretion column. More specifically, to reconcile the observed spin evolution of \swjs with the updated Gaia DR3 distance, the radiative efficiency should be lower by a factor of  $\sim(7/5.2)^2\sim 1.8$ (assuming standard NS parameters). State-of-the-art physical models about the emission of the accretion column generally assume that all gravitational energy is transformed to radiation \citep[e.g.][]{2016ApJ...831..194W,2017ApJ...835..130W,2017ApJ...835..129W}. As new fitting strategies are implemented into these models \citep{2021A&A...656A.105T} one could further test the radiative efficiency in the accretion column in systems like \swjs.

\subsection{Further application}
Our approach demonstrates that self-consistent modelling of the intrinsic and orbital spin-up of the system is essential for major outbursts. Coupling the torque models with a Bayesian interface delivers much more realistic uncertainties. Moreover, implementation of nested sampling may allow fitting data while using a wide parameter space for priors enabling better investigation of degeneracies and possible multimodal solutions. 
The Bayesian modelling is also useful for exploring orbital modulation in systems with low quality of data monitored with \swift/XRT or \nicer, as seen in a recent application we made in SXP\,15.6 \citep{2022A&A...664A.194V}.
In terms of the physical problem, inclusion of other torque models and extra terms would be the next step so the code can be applied to a wider family of accreting XRPs.
Finally, we plan to build upon our current code, and provide a user friendly version to the community with parallelization capabilities.

\section{Conclusion}\label{sec:conclusion}
We have used a nested sampling algorithm for Bayesian Parameter Estimation to study the spin evolution of nearby super-Eddington accreting pulsars. By coupling torque and orbital models for systems with well determined distance we were able to simultaneously estimate the orbital parameters and the magnetic field of the NS. A similar application to \swjs, the closest known PULX, revealed a transition that may be quantitatively linked to changes in the accretion disc structure close to the Eddington luminosity. According to the most recent Gaia parallax measurements \swjs seems to be closer than previously thought. The study of the NS spin up demonstrates that typical NS parameters cannot be used to explain the NS spin evolution using the updated distance. A possible solution is to assume a low-mass NS ($M\approx 1.1 M_{\odot}$) or assume a lower accretion column radiative efficiency (by a factor of 2) than typically assumed.  


\section*{Data availability}

X-ray data are available through the High Energy Astrophysics Science Archive Research Center: \url{heasarc.gsfc.nasa.gov}. 
\swift/BAT data are available through \swift transient monitoring project: \url{https://swift.gsfc.nasa.gov/results/transients/weak/}.
\fermi/GBM data are available through the GBM Accreting Pulsars project: \\ \url{https://gammaray.msfc.nasa.gov/gbm/science/pulsars.html}.

\section*{Acknowledgements}
We would like to thank the anonymous referee for a constructive report that helped to improve the manuscript.
%
Project was supported by Fermi Guest Investigator grant \#80NSSC20K1560.
MP acknowledges support from the MERAC Fondation through the project THRILL.
The project was supported by the Hellenic Foundation for Research and Innovation (H.F.R.I.) through the projects UNTRAPHOB (Project ID 3013) and ASTRAPE (Project ID 7802).
%
This research made use of 
Python v3.7.3, Astropy,\footnote{\url{http://www.astropy.org}} a community-developed core Python package for Astronomy \citep{2013A&A...558A..33A, 2018AJ....156..123A}, and {\sc ultranest} software package,\footnote{\url{https://johannesbuchner.github.io/UltraNest/}} for model-to-data comparison using nested sampling \citep{2021JOSS....6.3001B}. 




\bibliographystyle{mnras}
\bibliography{general}



\begin{appendix} 

\section{Corner plots and tables}\label{ap: cornerplots}

\begin{figure*}
 \includegraphics[width=\textwidth]{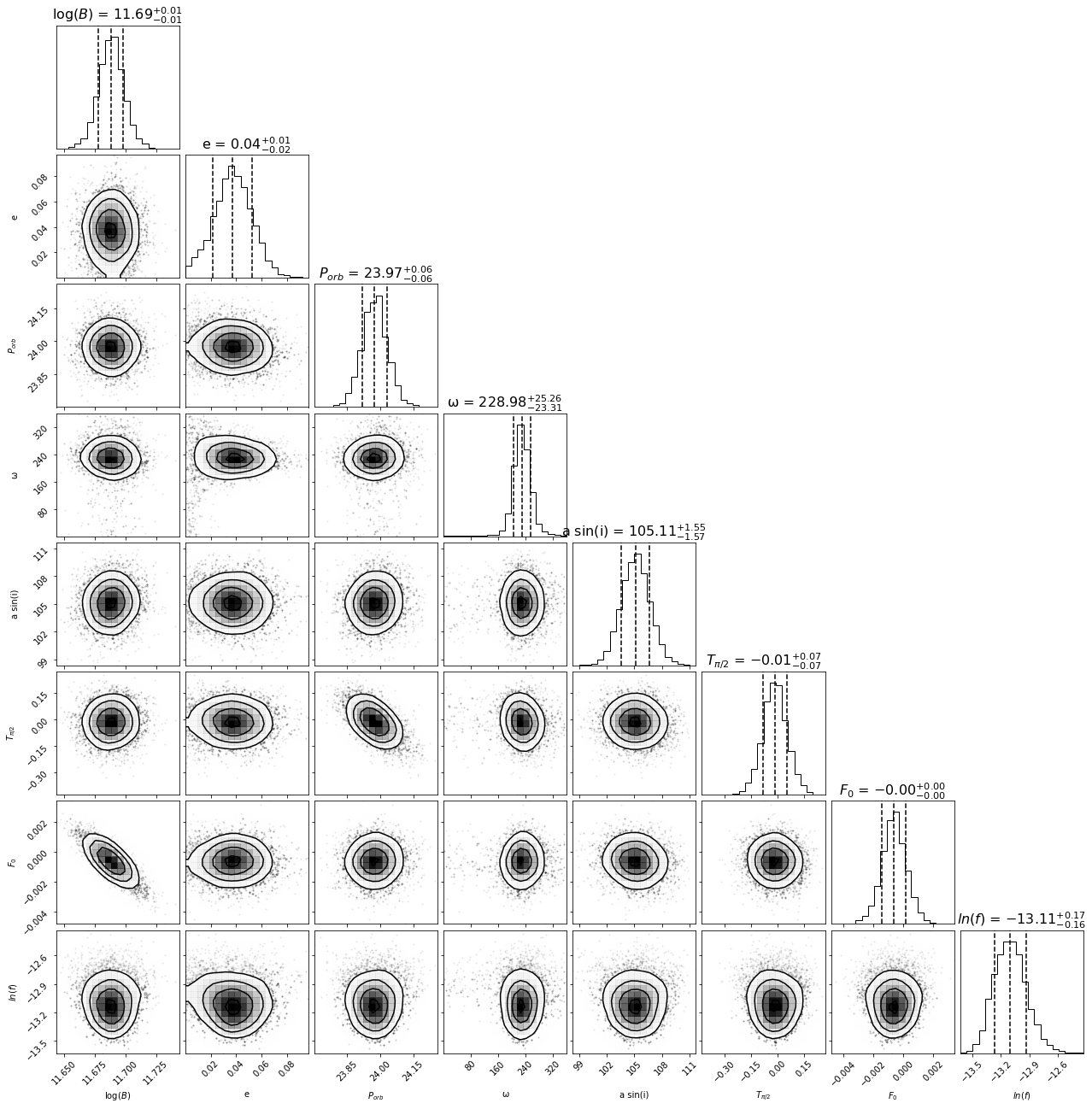}
    \caption{Corner plot for \lxps using the \protect\citetalias{1979ApJ...234..296G} model. We plot the logarithm of the magnetic field strength $B$ in G. The eccentricity ($e$), orbital period in days, the longitude of periastron in degrees ($\omega$) and the projected semi-major axis in light-sec ($a \, \sin i$). For clarity $T_{\rm \pi/2}$ is given relative to a reference MJD of 56666.91, while the reference frequency ($F_0$) is given relative to 124.3927 mHz. Finally,  $\ln(f)$ is the systematic scatter that is used to estimate the excess variance of the model. 
    }\label{fig:RXJ0520corner}
\end{figure*}

\begin{figure*}
 \includegraphics[width=\textwidth]{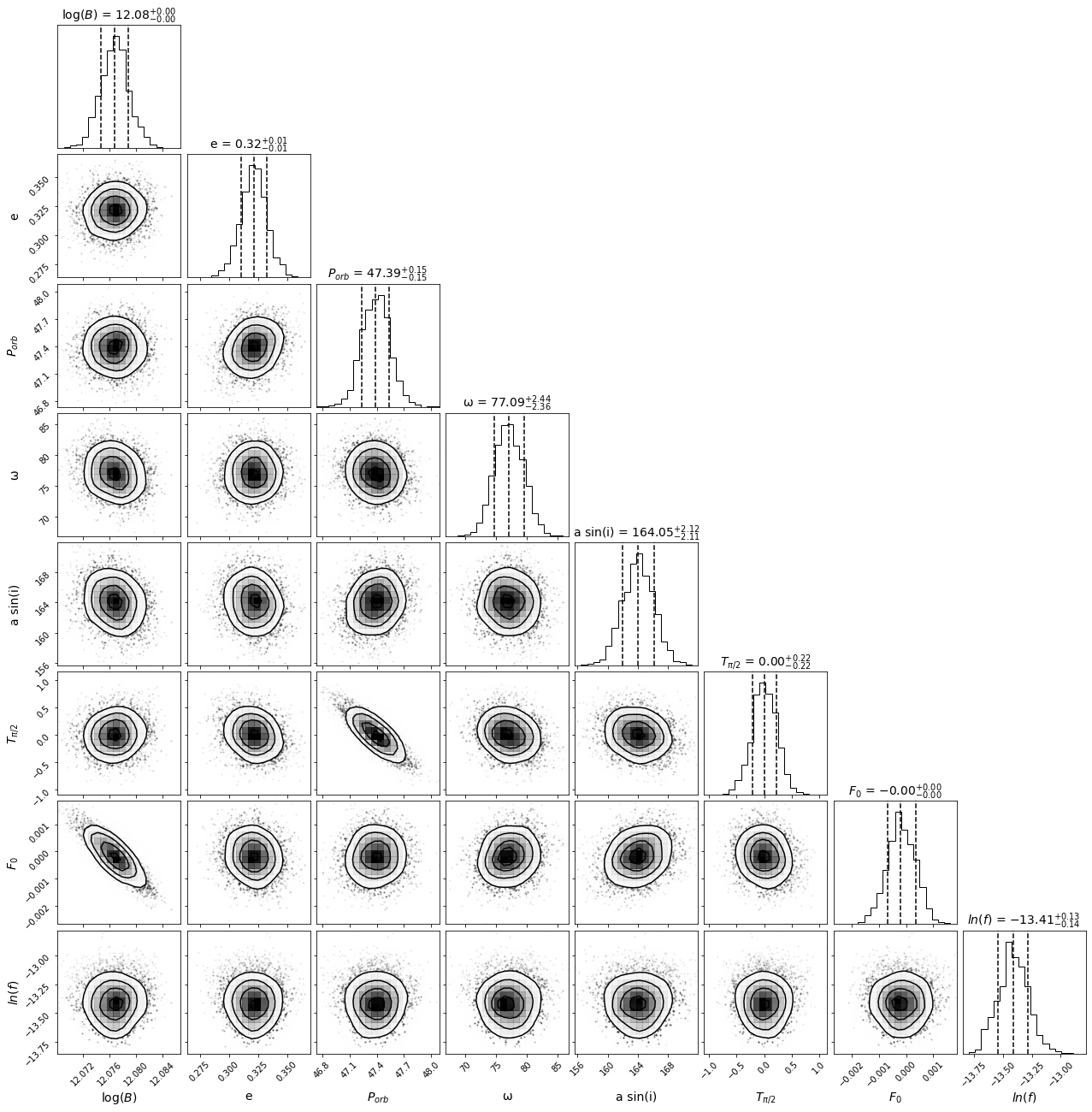}
    \caption{Corner plot for \rxjs using the GBM pulsed flux as a proxy of accretion rate and the \protect\citetalias{1995ApJ...449L.153W} model. We plot the logarithm of the magnetic field strength $B$ in G. The eccentricity ($e$), orbital period in days ($P_{\rm orb}$), the longitude of periastron in degrees (Per) and the projected semi-major axis in light-sec ($a \, \sin i$). For clarity, $T_{\rm \pi/2}$ is given relative to a reference MJD of 58793.32, while the reference frequency ($F_0$) is given relative to 107.4909 mHz. Finally,  $\ln(f)$ is the systematic scatter that is used to estimate the excess variance of the model.
    }\label{fig:RXJ0209_corner_Wang}
\end{figure*}

\begin{figure*}
 \includegraphics[width=\textwidth]{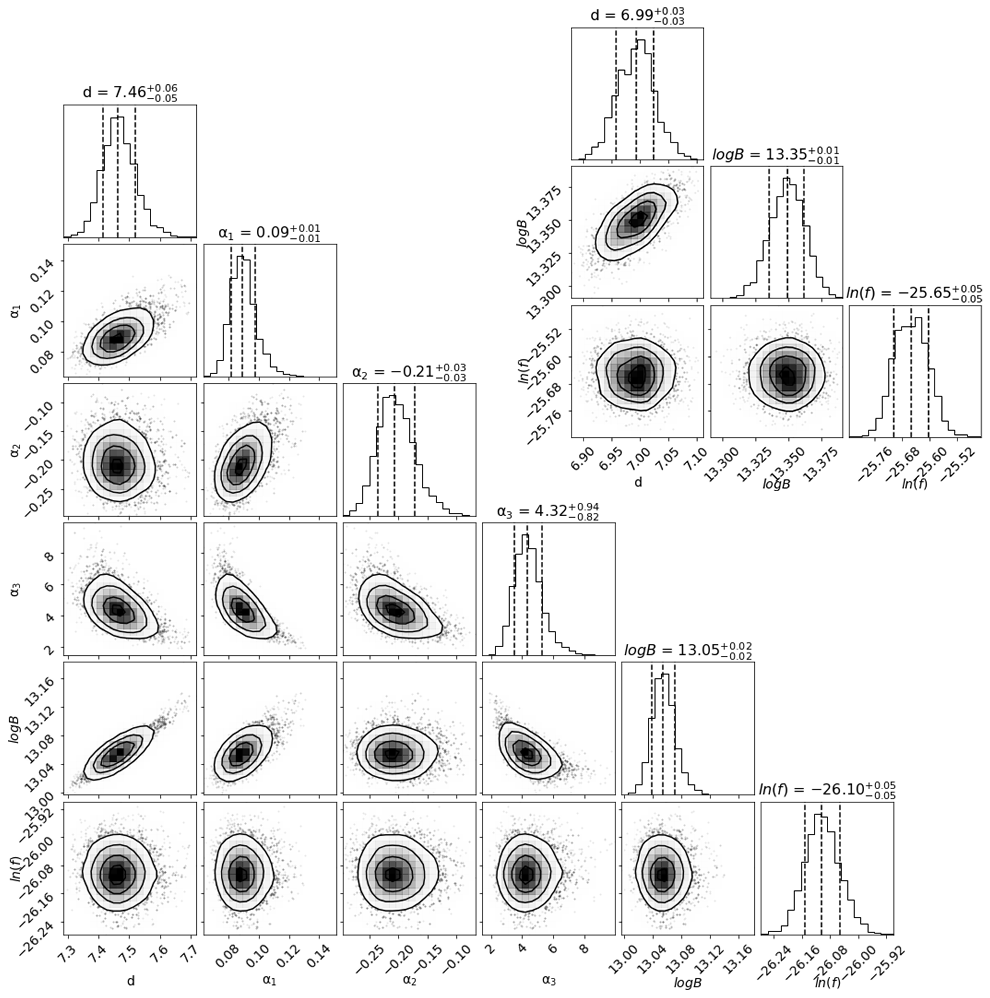}
    \caption{Corner plot for \swjs using the \swift/BAT count rates as a proxy of accretion rate, the \citetalias{2014MNRAS.437.3664H} model, our $\xi(\dot{m})$ model and fitting to the frequency derivative instead of the frequency. Instead of $a_{\rm 0}$ we plot the magnetic field logarithm $\log(B)$ directly, in G. Finally,  $\ln(f)$ is the systematic scatter that is used to estimate the excess variance of the model. The upper right contour is calculated for the standard $\xi = 0.5$ approach.}
    \label{fig:SWIFTJ_2_corner}
\end{figure*}

\begin{figure*}
 \includegraphics[width=\textwidth]{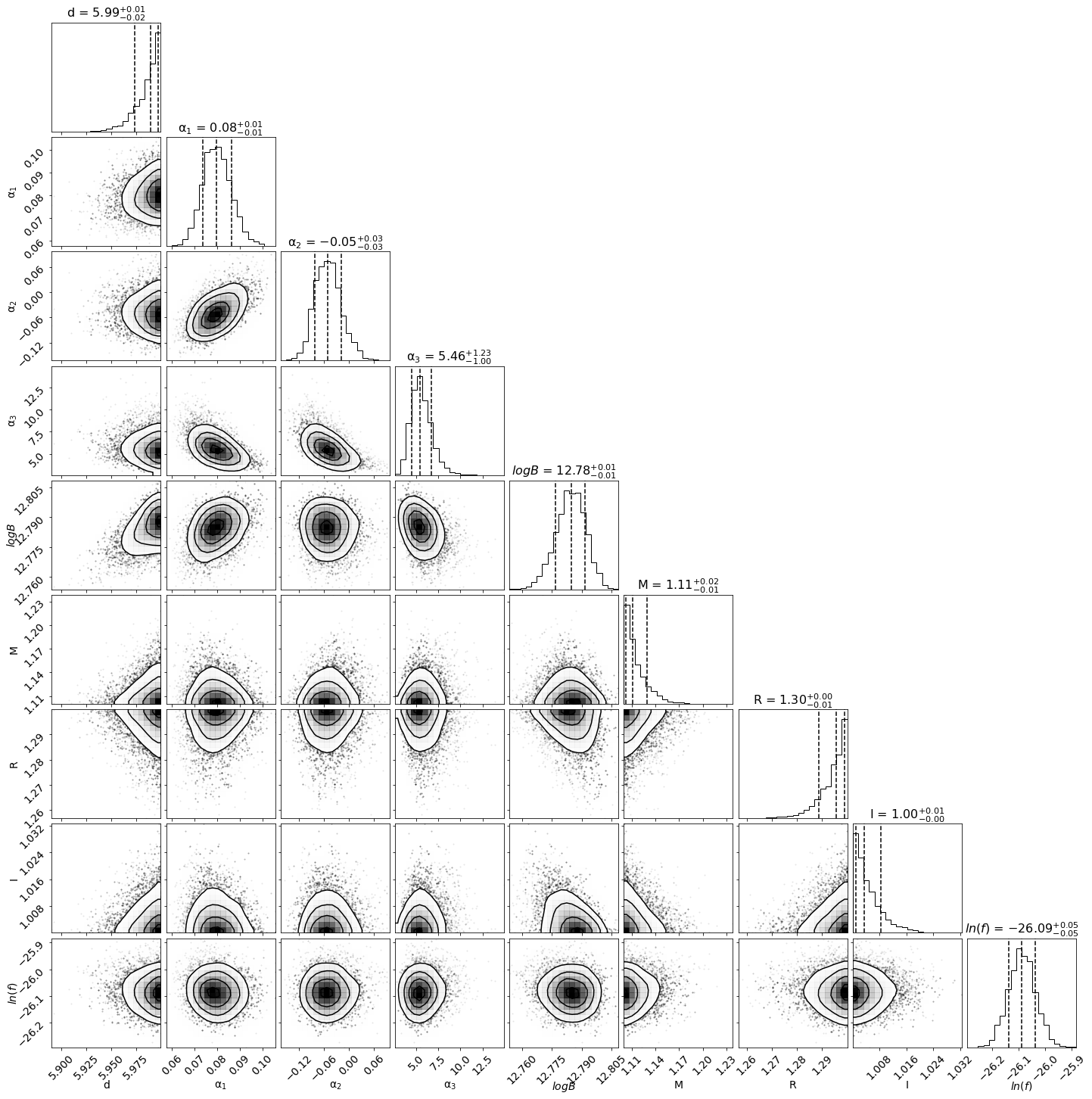}
    \caption{Corner plot for \swjs using the \swift/BAT count rates as a proxy of accretion rate, the \citetalias{2014MNRAS.437.3664H} model, our $\xi(\dot{m})$ model, with the NS mass $M$, radius $R$ and moment of inertia $I$ ass free parameters and fitting to the frequency derivative instead of the frequency. Instead of $a_{\rm 0}$ we plot the magnetic field logarithm $\log(B)$ directly, in G. The mass $M$ is given in $M_{\odot}$, the radius $R$ in $10^{6}{\rm\:cm}$ and the moment of inertia $I$ in $10^{45}{\rm\:g\:cm^{2}}$. Finally,  $\ln(f)$ is the systematic scatter that is used to estimate the excess variance of the model.}
    \label{fig:SWIFTJ_MIR_corner}
\end{figure*}

\begin{figure*}
 \includegraphics[width=\textwidth]{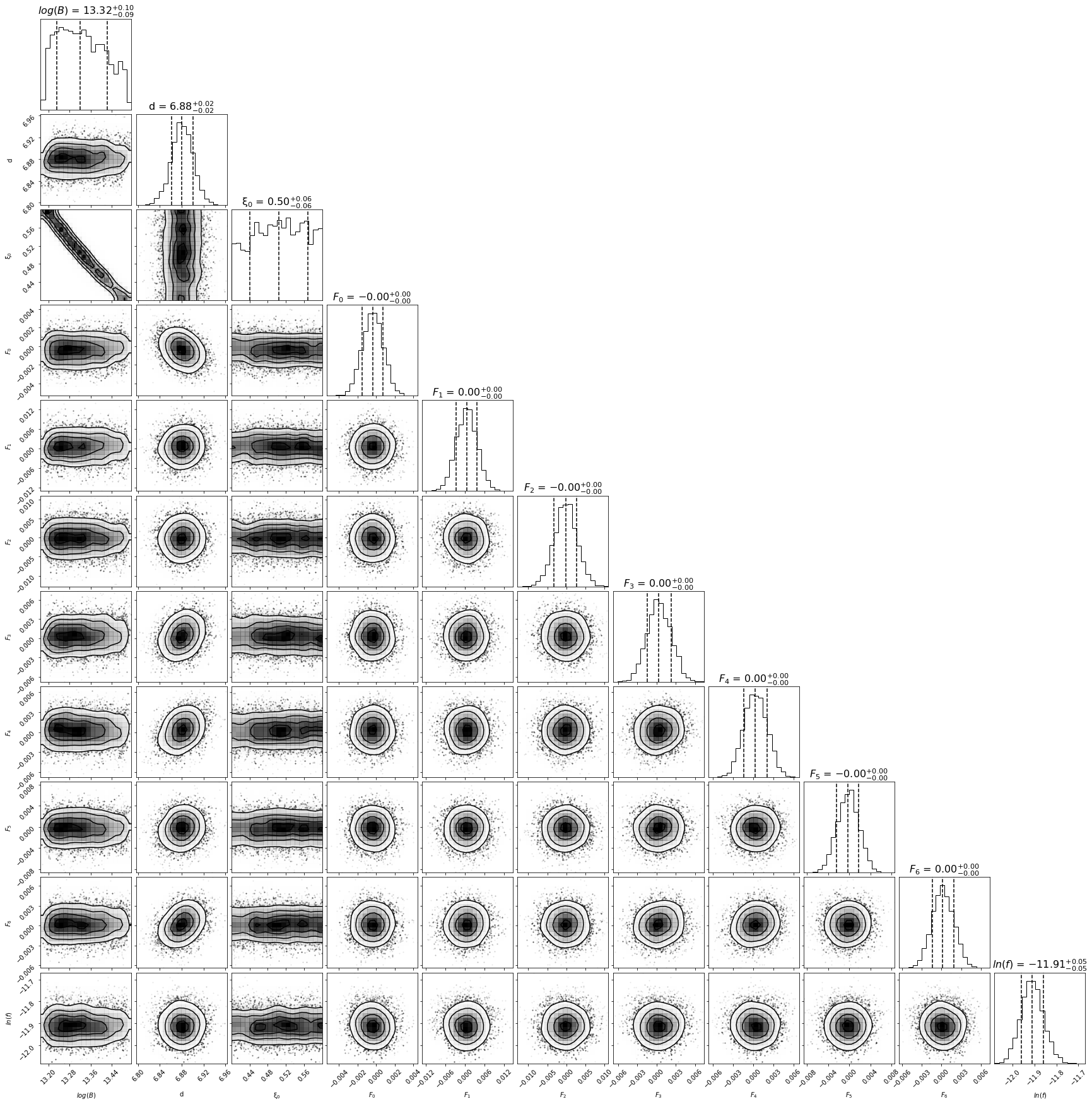}
    \caption{ Corner plot for \swjs using the \swift/BAT count rates as a proxy of accretion rate, the \citetalias{2014MNRAS.437.3664H} model and the standard $\xi =$const. approach. We plot the logarithm of the magnetic field strength $B$ in G. For clarity the reference frequencies ($F_{\rm 0}$ to $F_{\rm 6}$) are given relative to the result of the fit displayed in table \ref{tab:SWIFTresults}. Finally,  $\ln(f)$ is the systematic scatter that is used to estimate the excess variance of the model. 
    }\label{fig:SWIFTJ0243_corner_const_xi}
\end{figure*}

\begin{table*}
\caption{Results of modelling \swjs in the ${\nu}(t)$ parameter space.}
\begin{threeparttable}
\begin{tabular*}{\textwidth}[t]{p{0.2\textwidth}p{0.2\textwidth}p{0.2\textwidth}p{0.2\textwidth}p{0.11\textwidth}}
\hline\noalign{\smallskip} 
 Params\tnote{$\dagger$} & Reference MJD & $\xi =$const. &   $\xi(\dot{m})$ model & units\\
\hline\hline\noalign{\smallskip} 
\multicolumn{5}{l}{\bf{Torque model Parameters}}\\
\noalign{\smallskip} 	
         $\log{B}$  & -- & 13.33$\pm$0.09&   13.143$\pm$0.019\tnote{*} & G\\
         $v_{\rm 0}$ & $58027.5$ & 101.4850$\pm$0.0011&  101.4826$\pm$0.0004 & mHz\\
         $v_{\rm 1}$ & $58183.5$ & 102.136$\pm$0.003&   102.1339$\pm$0.0010 & mHz\\
         $v_{\rm 2}$ & $58239.6$ & 102.126$\pm$0.003&   102.1244$\pm$0.0009 & mHz\\
         $v_{\rm 3}$ & $58269.5$ & 102.1257$\pm$0.0019&   102.1212$\pm$0.0007 & mHz\\
         $v_{\rm 4}$ & $58335.5$ & 102.1137$\pm$0.0017&   102.1074$\pm$0.0007 & mHz\\
         $v_{\rm 5}$ & $58401.5$ & 102.1013$\pm$0.0021&   102.0989$\pm$0.0007 & mHz\\
         $v_{\rm 6}$ & $58440.6$ & 102.0967$\pm$0.0017&   102.0899$\pm$0.0006 & mHz\\
 		 $\xi_{\rm 0}$ & -- & 0.50$\pm$0.06&     -- & --\\
 		 $a_{\rm 1}$ & -- & -- &     0.104$\pm$0.006 & --\\
 		 $a_{\rm 2}$ &-- & -- &     -0.285$\pm$0.009 & --\\
 		 $a_{\rm 3}$ &-- & -- &     3.7$\pm$ 0.5 & --\\
 		 $a_{\rm 0}$ &-- & -- &      3.650$\pm$0.011& --\\
\hline\noalign{\smallskip} 	
\multicolumn{5}{l}{\bf{Other Parameters}}\\\noalign{\smallskip}
     $d$  &-- & 6.881 $\pm$ 0.020  &   7.75$\pm$0.07 & kpc\\
 	$\ln{f}$ &--& -11.91$\pm$0.05&    -13.08 $\pm$ 0.05& --\\
\hline\noalign{\smallskip}
\multicolumn{5}{l}{\bf{Evidence}}\\\noalign{\smallskip}
 	$\ln{Z}$ &--& 2408.3$\pm$0.6 &   2642.1 $\pm$ 0.5& --\\
\hline\noalign{\smallskip}  
\end{tabular*}
\begin{tablenotes}\footnotesize
\item[$\dagger$] Reported values of the fitted parameters and their uncertainties are estimated from the mean and standard eviation of the constructed posterior samples with {\sc ultranest}.
\item[*] Inferred from $a_{\rm 0}$.
\end{tablenotes}
\end{threeparttable}\label{tab:SWIFTresults}
\end{table*}

\end{appendix}


\bsp	
\label{lastpage}
\end{document}